\documentclass[
               secnumarabic,
               amssymb, 
               nobibnotes, 
               aps, 
               prd, 
               longbibliography,
               amsmath]{revtex4-1}

\usepackage{graphicx}
\usepackage{amsmath}
\usepackage{bm}

\begin{document}

\preprint{}

\title{On the wake meandering of a model wind turbine operating in two different regimes}



\author{Daniel Foti}
\affiliation{Department of Mechanical Engineering, St. Anthony Falls Laboratory, University of Minnesota, Minneapolis, Minnesota 55455, USA}

\author{Xiaolei Yang}
\affiliation{Department of Civil Engineering, Department of Mechanical Engineering, College of Engineering and Applied Sciences, Stony Brook University, Stony Brook, New York 11794, USA}
\author{Filippo Campagnolo}
\affiliation{Wind Energy Institute, Technische Universit\"{a}t M\"{u}nchen, Garching bei M\"{u}nchen, Germany}

\author{David Maniaci}
\affiliation{Sandia National Laboratories, Albuquerque, New Mexico 87185, USA}

\author{Fotis Sotiropoulos}
\affiliation{Department of Civil Engineering, College of Engineering and Applied Sciences, Stony Brook University, Stony Brook, New York 11794, USA}

\begin{abstract}
\textbf{The flows behind a model wind turbine under two different turbine operating regimes 
(Region 2 for turbine operating at optimal condition with the maximum power coefficient and 1.4 degrees of pitch angle, and Region 3 for turbine operating at sub-optimal condition with a lower power coefficient and 7 degrees of pitch angle) 
are investigated using wind tunnel experiments and large-eddy simulations (LES).  Measurements from the model wind turbine experiment reveal that the power coefficient and turbine wake are affected by the operating regime.    
Simulations employing a new class of actuator surface methods which parameterize both the turbine blades and nacelle with and without a nacelle model are carried out for each operating condition to study the influence of the operating regime and nacelle on the formation of the hub vortex and wake meandering.  Flow field statistics and energy spectra of the simulated wakes are in good agreement with the measurements.   
For simulations with a nacelle model, the mean flow field is composed of an outer wake, caused by energy extraction from the incoming wind by turbine blades, and an inner wake directly behind the nacelle, while for the simulations without a nacelle model, the central region of the wake is occupied by a jet.  The simulations with the nacelle model reveal an unstable helical hub vortex expanding outwards towards the outer wake; while the simulations without a nacelle model show a stable and columnar hub vortex.  Because of the different interactions of the inner region of the wake with the outer region of the wake, a region with higher turbulence intensity is observed in the tip shear layer for the simulation with a nacelle model. The hub vortex for the turbine operating in Region 3 remains in a tight helical spiral and intercepts the outer wake a few diameters further downstream than for the turbine operating in Region 2.  Wake meandering, a low frequency large-scale motion of the wake, commences in the region of high turbulence intensity for all simulations with and without a nacelle model indicating that neither a nacelle model nor an unstable hub vortex is a necessary requirement for the existence of wake meandering.  However, further analysis of the wake meandering and instantaneous flow field using a filtering technique and dynamic mode decomposition show that the unstable hub vortex energizes the wake meandering.  The turbine operating regime affects the shape and expansion of the hub vortex altering the location of the onset of the wake meandering and wake meander oscillating intensity.  Most importantly, the unstable hub vortex promotes a high amplitude energetic meandering which cannot be predicted without a nacelle model. } 
\end{abstract}

\maketitle
%
%
%
%
%
%
\section{Introduction}
Power losses due to the turbine wake are reported as high as 20\% \citep{barthelmie2008flow}. In addition to other conditions of ambient flow and wind turbines, e.g. turbulence intensity of incoming flow, and thrust and power coefficients of a wind turbine, the velocity deficit and turbulence intensity of the turbine wake are also significantly affected by a low frequency large-scale coherent oscillation, the so-called wake meandering. However, the wake meandering mechanism is still far from fully understood \citep{okulov2014regular, kang2014onset}, and low-order models \citep{larsen2008wake,yang2012computational} cannot take into account the wake meandering effects accurately. This poses a significant difficulty for wind farm control and optimization. In this study, we attempt to elucidate details on the dynamics of turbine wake meandering for different turbine operating conditions and assess the corresponding nacelle effects. \\
\indent Directly behind the turbine a system of helical vortices dominates the flow.  As first proposed by \citet{joukowski1912vortex} for a $N$-bladed propeller, the wake consists of N helical tip vortices, each with circulation $\Gamma$, shed from the tip of each turbine blade and an $N\Gamma$ single counter rotating hub vortex oriented along centerline in the streamwise direction.   If the circulation along the blade is not uniform, short-lived trailing vortices also shed from the trailing edge of the blade. Tip vortices have been characterized by several experimental techniques (\cite{chamorro2009wind, hu2012dynamic, sherry2013interaction, hong2014natural}) and numerical studies (\cite{ivanell2009analysis, ivanell2010stability, troldborg2007actuator}). These studies have shown that the tip vortices convect downstream due to the relatively high speed flow near the blade tip and eventually breakdown, a process which depends on many factors, such as the turbulence in the incoming flow, rotational speed, geometry of the turbine blade and the interactions of helical vortices. It was shown in the theoretical work of \citet{widnall1972stability} on helical vortex instability that the tip vortex instability includes both short-wave and long-wave instabilities but is mainly caused by mutual induction because of the pitch of the helical structure. \citet{felli2011mechanisms} were able to observe all three instability modes for a propeller in a water tunnel, while \citet{sarmast2014mutual} showed the prevalence of the mutual induction in tip vortices using mode decomposition on tip vortices obtained from a numerical simulation.
Recently, \citet{yang2016coherent} employed large-eddy simulation (LES) in conjunction with large-scale snow particle image velocimetry (PIV) experiments to investigate the complexity of the coherent structures in the tip shear layer of a utility scale turbine and uncovered a new instability mode of the tip vortices.   
The hub vortex has been studied less extensively than the tip vortices.  \citet{iungo2013linear} used linear stability analysis on a model turbine and showed that the hub vortex is also unstable.  \citet{felli2011mechanisms} visualized the hub vortex and witnessed its breakdown with relation to the tip vortices.  \citet{okulov2007stability} performed theoretical stability analysis and concluded that the system of tip vortices and a hub vortex is unconditionally unstable. 
In recent geometry-resolving simulations of a hydrokinetic turbine \citep{kang2014onset} and a model wind turbine \citep{foti2016wake} using LES and the curvilinear immersed boundary method, it was shown that the helical hub vortex forms behind the nacelle and expands in an inner wake (a strong wake at the centerline of the turbine wake) formed by the nacelle.  The inner wake recovers and expands quickly allowing the hub vortex to interact with the outer wake. The work of \citet{kang2014onset} and \citet{foti2016wake} further showed that such interaction of the hub vortex with the outer wake augments the intensity of wake meandering at far wake locations. \\
\indent \citet{medici2008measurements} first explored experimental results of meandering in the far wake of the turbine. The oscillations were attributed to bluff-body vortex shedding effects. Similar low frequency oscillations are recorded in \citet{chamorro2013interaction}, \citet{okulov2014regular}, and \citet{howard2015statistics}.  In these studies, a similar value of the Strouhal number, the non-dimensional frequency normalized by the rotor diameter and incoming velocity at hub height was observed for various turbines with different operating conditions, which is about $0.3$.  Statistics of the amplitude and curvature of the wake meandering phenomena for a model wind turbine under different operating conditions were examined in \citet{howard2015statistics} by using a filtering technique on the data obtained from PIV measurements.   \\
\indent Numerical investigations, especially using LES, have been used recently to investigate the turbine wake.  Due to the high spatial resolution requirement for the geometry-resolving simulations using immersed boundary methods, as found in \citet{kang2014onset} and \citet{foti2016wake}, turbine modeling in the form of actuator-based models of actuator disks or actuator lines have been used exceedingly. The actuator disk, a model consisting of a porous disk concept as in \citet{glauert1935airplane, yang2012computational, calaf2010large, porte2011large} and actuator lines modeling the blades as rotating lines with distributed lift and drag forces \citep{so̸rensen2002numerical, ivanell2010stability, yang2015large}, are commonly employed. 
Previous numerical (e.g. \citet{yang2012computational, calaf2010large, yang2015large}) and experimental (e.g. \citet{espana2011spatial}) results showed that both actuator disk models and actuator line models are able to produce large-scale motion of the wake consistent with wake meandering.  However, actuator-based models without accounting for the influence of the nacelle may not be able to capture the hub vortex expansion and its interaction with the outer wake and thus cannot accurately describe the effects of nacelle on the dynamics of wake meandering.
\citet{kang2014onset} simulated the flow past a hydrokinetic turbine using geometry-resolving immersed boundary method, actuator disk model and actuator line model. The study found that the hub vortex from the actuator-type simulations remains columnar without significant interaction with the outer wake resulting in significantly less turbulence kinetic energy in the far wake.  
In a recent work by \citet{yang2017actuator}, a new class of actuator surface models for turbine blades and nacelle was developed and validated with the hydrokinetic turbine \citep{kang2014onset} and MEXICO (Model Experiments in Controlled Conditions) turbine cases with overall good agreement. Being able to capture the interaction of the hub vortex with the outer wake on a relatively coarse grid, the actuator models for turbine blades and nacelle provide a computationally efficient approach to investigate the effects of nacelle and different turbine and ambient flow conditions on wake meandering. \\
\indent In this work, we employ both wind tunnel experiments and LES with the new class of actuator surface models developed by \citet{yang2017actuator} to investigate: i) the influence of turbine operating conditions; and ii) the effects of the nacelle on wake meandering of a model turbine.
In the literature, there are several wind tunnel experiments investigating turbine wakes under different operating conditions, such as experiments of the ``Blind Test'' turbine of diameter 0.9 m and optimal tip speed ratio 6 at Norwegian University of Science and Technology \citep{krogstad2012performance, eriksen2012experimental, adaramola2011experimental, pierella2014blind}, and experiments of the MEXICO turbine of rotor diameter 4.5 m and optimal tip speed ratio 6.7 at the German-Dutch wind tunnel~\citep{snel2007mexico, shen2012actuator, nilsson2015validation}. Very few experiments addressed the influence of turbine operating conditions on wake meandering.
The effects of turbine operating condition on wake meandering were studied by \citet{howard2015statistics}. However, we note that the turbine employed in this paper, with diameter 1.1 meters, has more reasonable power coefficients (the maximum $C_P$ is 0.45) as compared with that (the maximum $C_P$ is 0.16) of the miniature model wind turbine with diameter 0.128 meters employed in \citet{howard2015statistics}.  We consider, therefore, that the wake of the turbine employed in this work closely resembles the wake state of utility-scale turbines. 
The model turbine utilized herein for wind tunnel experiments is intricately designed such that individual blades can be pitched similar to variable speed utility-scale turbines which employ blade pitching to control power output based on the operating conditions.
Like utility-scale turbines, the model wind turbine exhibits three main turbine operating regions \citep{pao2009tutorial}. 
Region 1 is a low wind speed regime when the turbine is not run. In Region 2, the turbine operates at its optimal condition with maximum power coefficient and blade pitch angle of 1 degree; while in Region 3,  the turbine operates with a relatively low power coefficient and a relatively high blade pitch angle of 7 degrees. 
To investigate the energetic coherent structures of wake meandering, the meander filtering technique proposed by \citet{howard2015statistics}, and applied by \citet{foti2016wake}, to the miniature model wind turbine with diameter 0.128 meters, and the dynamic mode decomposition technique \citep{schmid2010dynamic} are employed to analyze the time series of the computed three-dimensional flow fields. \\
\indent This paper is organized beginning with section \ref{sec:exp_setup}, a brief summary of the wind turbine design and experimental setup which provide motivation and validation for simulations.  The numerical methods of the LES and turbine modeling in section \ref{sec:numerical}. Section \ref{sec:cases} details of the selected test cases and computational setup.  The results and analysis are shown in section \ref{sec:results}.   Finally, we have final discussions of our results and our conclusion in section \ref{sec:conclusion}.
%
%
%
%
%
%
\section{Experimental setup}\label{sec:exp_setup}
\indent Wind tunnel experiments and numerical simulations play a key role in wind turbine design, performance, and optimization.  While the wind tunnel experiments cannot reproduce the utility-scale conditions, they can provide key insights and ensure that numerical simulations are reliable through proper validation.  The present experiments supply sufficient detail in terms of measurements of the velocity flow field statistics and energy spectra in the wake of the turbine, turbine power output, and a robust description of the model and wind tunnel environment that allow us to validate numerical simulations of the model turbine in different operating regimes.  Below we detail the wind tunnel experiment, wind turbine model and the turbine power output as a function of blade pitch and tip speed ratio which delineates the turbine operating regimes. \\   
\indent Testing was performed in the closed-return wind tunnel of the Politecnico di Milano (see \citet{campagnolo2013wind}). 
It a boundary layer chamber primarily used for civil, environmental and wind power engineering applications. Within this test section, whose cross sectional area is $13.84\:\textrm{m}\times 3.84$~m with a length of 36~m, wind speeds up to 14~m/sec can be generated by means of 14 fans, each one consuming up to 100~KW.\\
\indent This tunnel size allows for the testing of relatively large models with low blockage effects, while atmospheric boundary layer (ABL) conditions can be simulated by the use of turbulence generators such as spires placed at the chamber inlet.  A turntable, whose diameter is 13~m, allows for the complete experimental setup to be yawed with respect to the wind tunnel axis, in order to simulate the effect of wind direction changes on the entire setup. 
%
%
\subsection{Wind turbine model: general layout}
Tests were conducted with a scaled wind turbine model with a diameter $D = 1.1 \: \textrm{m}$ and  height $H = 0.8 \: \textrm{m}$, in the following named \texttt{G1} (for \underline{G}eneric wind turbine, \underline{1}~m diameter rotor). The model was designed to satisfy several specific design requirements: i.) a realistic energy conversion process, which means reasonable aerodynamic loads and damping when compared to those of full-scale wind turbines, as well as wakes of realistic geometry, velocity deficit and turbulence intensity, ii.) collective blade pitch and torque control, as well as yaw setting realized by properly misaligning the tower base with respect to the wind tunnel axis, in order to enable the testing of control strategies, and iii.) a sufficient onboard sensorization of the machine, including measures of rotor azimuth, main shaft torque, rotor speed and tower base loads with good accuracy.\\
\begin{figure}
    \begin{center}
    \includegraphics[width=.75\columnwidth]{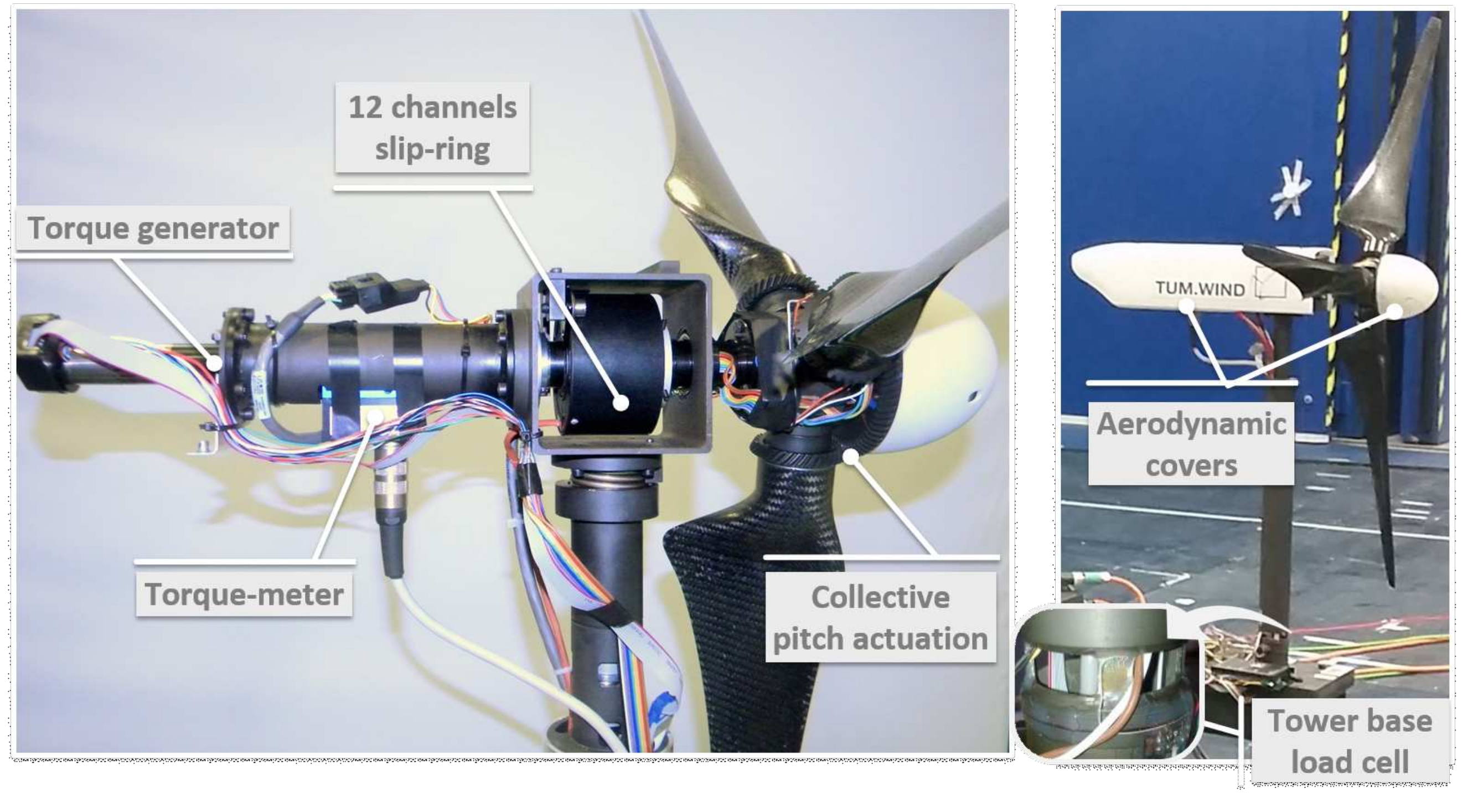}
    \caption{\label{figure: G1 rotor-nacelle assembly} Overall view of the \texttt{G1} model.}
   \end{center}
\end{figure}
\indent The \texttt{G1}, whose rated rotor speed is equal to 850\,rpm (clockwise rotation), is equipped with three blades mounted on the hub with two bearings, in order to limit flapwise or edgewise free-play. The collective pitch angle can be varied by means of three conical spiral pinions fixed at the blade roots, in turn moved by a driving wheel. The latter, mounted on two bearings held by the rotating hub, is connected, by means of a flexible joint, to a Maxon \texttt{EC-16 60W} brushless motor equipped with a \texttt{GP22C-128:1} precision gearhead and \texttt{MR-128CPT} relative encoder. The motor is housed in the hollow shaft and is commanded by an
electronic control board \texttt{EPOS2 24/2 DC} housed in the hub spinner.
Electrical signals from and to the pitch control board are transmitted by a through-bore 12-channels slip ring located within the rectangular carrying box holding the main shaft. \\
\indent A \texttt{LORENZ MESSTECHNIK DR2112-R} torque sensor 1\,Nm 0.2\%, located after the two shaft bearings, allows for the measurement of the torque provided by a Maxon \texttt{EC-4pole22 90W} brushless motor equipped with a \texttt{GP22HP-14:1} precision gearhead and \texttt{ENC HEDL 5540 500IMP} tacho. The motor is located in the rear part of the nacelle and is operated as a generator by using an \texttt{ESCON 50/5 4-Q} servocontroller. An optical encoder, located between the slip ring and the
rear shaft bearing, allows for the measurement of the rotor azimuth. \\
\indent The tower is designed so that the first fore-aft and side-side natural frequencies of the nacelle-tower group are properly placed with respect to the harmonic per-rev excitations. At its base, strain gages are glued to four CNC-machined small bridges, which were properly sized in order to have sufficiently large strains, in turn required to get accurate outputs from the strain gages. Two electronic boards provide for the power supply and adequate conditioning of this custom-made load cell. Calibration of the cell was performed by using dead weights, stressing the tower with fore-aft and side-side bending moments. Finally, a full 2-by-2 sensitivity matrix is obtained by linear regression. 
Aerodynamic covers of the nacelle and hub ensure a satisfactory quality of the flow in the central rotor area. Fig.~\ref{figure: G1 rotor-nacelle assembly} highlights the main features of the \texttt{G1} model.
%
%
\subsection{Wind turbine model: rotor aerodynamics}
\begin{figure}[h]
   \begin{center}
       \includegraphics[width=\textwidth]{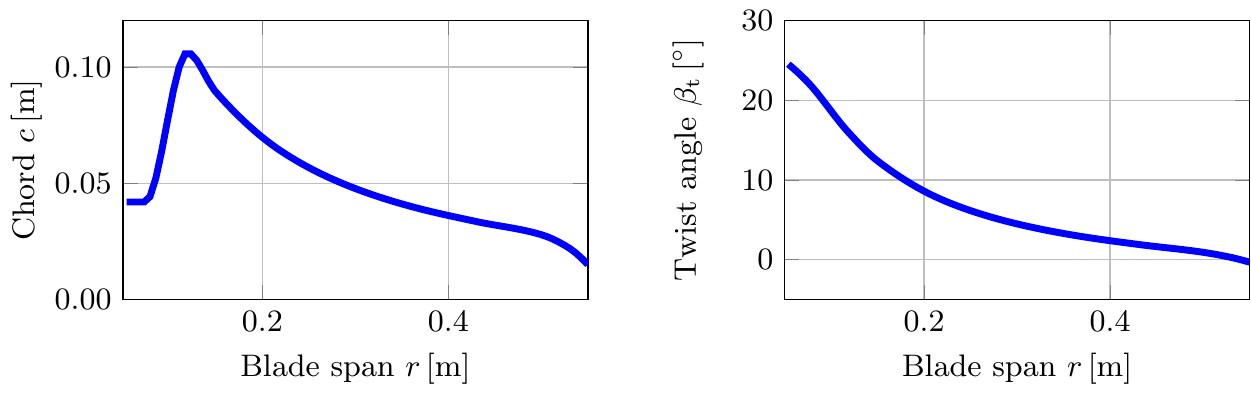}
       \caption{\label{fig:Blade_design}Blade chord and twist angle.}
   \end{center}
\end{figure}

\begin{figure}
   \begin{center}
     \includegraphics[width=\textwidth]{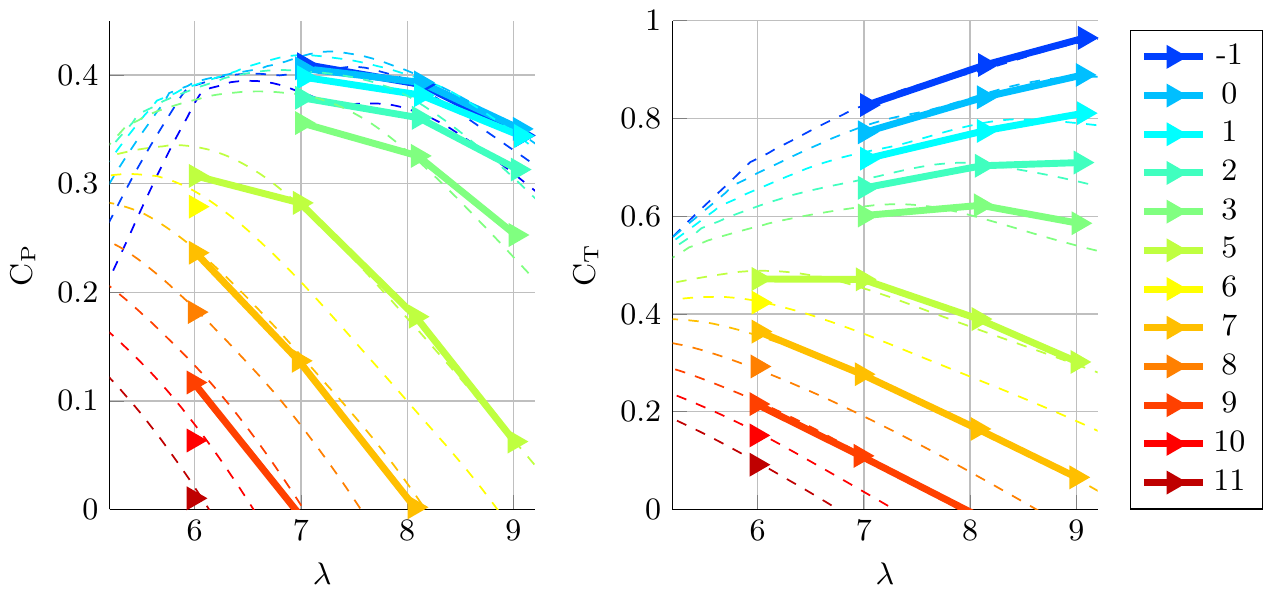}
     \caption{\label{figure: rotor performance} \texttt{G1} power (left) and thrust (right) experimental (solid lines) and BEM (dashed lines) coefficients, as function of TSR and blade pitch.}
   \end{center}
\end{figure}

Due to the small dimensions of the scaled wind turbine the low-Reynolds Number airfoil RG14~\citep{Airfoils} is chosen for the model wind turbine blades, which is designed in order to achieve a constant lift coefficient $C_\mathrm{l}$ along the blade span, as described in~\citet{Burton2001}. Blade chord and twist angle are shown in Fig.~\ref{fig:Blade_design}
The performance of the \texttt{G1} rotor is measured for different values of the airfoil Reynolds numbers and at several combinations of tip speed ratios (TSR) $\lambda$ and collective pitch settings $\beta$. Significant differences are noticed between the measured and theoretical Blade Element Momentum (BEM)-based aerodynamic performance computed using nominal polars, obtained by other authors from wind tunnel measurements or numerical simulations. To correct for this problem, an identification procedure~\citep{Cacciola2014} is used to calibrate the polars, leading to a satisfactory agreement as shown in Fig.~\ref{figure: rotor performance}.  
%
%
\subsection{Wind turbine model: control algorithms}
The \texttt{G1} model is controlled by a \texttt{M1 Bachmann} hard-real-time module. Similarly to what is done on real wind turbines, collective pitch-torque
control laws are implemented on and real-time executed by the control hardware. Sensor readings are used online to properly compute the desired pitch and torque
demands, which are in turn sent to the actuator control boards via analog or digital communication. \\
\indent Power control is based on the standard wind turbine control structure with two distinct regions, as described in~\citet{Bossanyi2000a}. At low wind speeds, when the wind turbine is operating in the Region 2, the main objective is the maximization of wind turbine power output. This is achieved by keeping the rotor blade pitch angle at a constant value, while the aerodynamic torque reference follows a quadratic function of rotor speed. 
On the other hand, in high winds,  when the wind turbine is operating in the Region 3, generator torque is kept at a constant value (defined by the power reference), while a PID pitch controller is used to track the rotor speed reference.  Transition between two different operating regions is achieved by a control logic that prevents pitch activity at low wind speeds and torque activity at high wind speeds. \\
\indent The friction acting on the main shaft and due to the bearings and the slip ring brushes is measured before testing, in order to have a reliable measurement of the aerodynamic power. This is obtained by rotating the sole rotor hub, i.e. without blades installed, at several speeds, from rated down to null in steps of 50\,rpm. The average measured generator torque is than stored in look-up tables as a function of rotor speed, and it is added during operation in real time to the generator torque reference, in order to allow for the tracking of the aerodynamic torque reference.
%
%
%
%
\section{Numerical methods}\label{sec:numerical}
\indent  The measurements of the flow field behind the wind turbine are used in part for validation of numerical methods.  For the present numerical simulations, we employ a large-eddy simulation (LES) solver capable of solving the three-dimensional, filtered continuity and momentum equations in generalized curvilinear coordinates \citep{ge2007numerical} with a hybrid staggered/non-staggered grid formulation \citep{gilmanov2005hybrid}.  The governing equations are discretized with three-point central finite differencing and integrated in time using an efficient fractional step method. In compact tensor notation (repeated indices imply summation) the continuity and momentum equations are as follows ($i,j=1,2,3$): 
\begin{equation}
J\frac{\partial U^{i}}{\partial \xi^{i}}=0,
\label{eqn:eq_continuity_general}
\end{equation}
\begin{align}
\frac{1}{J}\frac{\partial U^{i}}{\partial t}=& \frac{\xi _{l}^{i}}{J}\left( -%
\frac{\partial }{\partial \xi^{j}}({U^{j}u_{l}})+\frac{\mu}{\rho}%
\frac{\partial }{\partial \xi^{j}}\left(  \frac{g^{jk}}{J}\frac{%
\partial u_{l}}{\partial \xi^{k}}\right) -\frac{1}{\rho}\frac{\partial }{\partial \xi^{j}} \left(\frac{%
\xi _{l}^{j}p}{J} \right)-\frac{1}{\rho}\frac{\partial \tau _{lj}}{\partial
\xi^{j}} + F_l\right) ,
\label{eqn:eq_momentum_general}
\end{align}
where $\xi _{l}^{i}={\partial \xi^{i}}/{\partial x_{l}}$ are the transformation metrics, $J$ is the Jacobian of the geometric transformation, $u_{i}$ is the $i$th component of the velocity vector in Cartesian coordinates, $U^{i}$=${(\xi _{m}^{i}/J)u_{m}}$ is the contravariant volume flux, $g^{jk}=\xi _{l}^{j}\xi _{l}^{k}$ are the components of the contravariant metric tensor, $\rho $ is the density, $\mu $ is the dynamic viscosity, $p$ is the pressure, and $\tau_{ij}$ represents the anisotropic part of the subgrid-scale stress tensor.  A body force $F_l$ is used to account for the forces exerted by the actuator-based model.  A dynamic Smagorinsky model \citep{smagorinsky1963general} developed by \citet{germano1991dynamic} is used for closure of $\tau _{ij}$:
\begin{equation}
{\tau }_{ij}-\frac{1}{3}\tau _{kk}\delta _{ij}=-2\mu_{t}\widetilde{S}_{ij},
\label{eqn:LES_subgrid_eq}
\end{equation}
where the $\widetilde{(\cdot)}$ denotes the grid filtering operation, and $\widetilde{S}_{ij}$ is the filtered strain-rate tensor.
The eddy viscosity $\mu_{t}$ is given by
\begin{equation}
\mu_{t}=\rho C_{s}{\Delta}^{2}|\widetilde{S}|,
\label{eqn:LES_eddyviscosity_eq}
\end{equation}
where $C_{s}$ is the dynamically calculated Smagorinsky constant \citep{germano1991dynamic}, $\Delta$ is the filter size taken as the cubic root of the cell volume, and $|\widetilde{S}|= (2\widetilde{S}_{ij}\,\widetilde{S}_{ij})^{\frac{1}{2}}$.  In computing $C_{s}$, contraction of the Germano identity is carried out using the formulation for general curvilinear coordinates presented in \citet{armenio2000lagrangian}. A local averaging is then performed for the calculation of $C_{s}$ since there are no homogeneous directions in the present cases.   
%
%
\subsection{Actuator surface model for turbine blades and nacelle}\label{sec:actuator surface model}
\indent In the actuator surface model for the blade, the blade geometry is represented by a surface formed by the chord lines at every radial location of the blade. 
The forces are calculated in the same way as in the actuator line model using the blade element approach. 
The lift ($\bm{L}$) and drag ($\bm{D}$) at each radial location are calculated as follows:
\begin{equation}
\bm{L}=\frac{1}{2} \rho C_L c |\bm{V}_{rel}|^2\bm{n}_{L},
\end{equation}
and
\begin{equation}
\bm{D}=\frac{1}{2} \rho C_D c |\bm{V}_{rel}|^2\bm{n}_{D}, 
\end{equation}
where $C_L$ and $C_D$ are the lift and drag coefficients from a look-up table based on the value of angle of attack, $c$ is the chord, $\bm{V}_{rel}$ is the relative incoming velocity, and $\bm{n}_{L}$ and $\bm{n}_{D}$ are the unit vectors in the directions of lift and drag, respectively. The relative incoming velocity $\bm{V}_{rel}$ is computed by 
\begin{equation}
\bm{V}_{rel} (\bm{X}_{LE})= u_x(\bm{X}_{LE})\bm{e}_x+(u_\theta(\bm{X}_{LE})-\Omega r)\bm{e}_\theta
\end{equation}
where $\bm{X}_{LE}$ represents the leading edge coordinates of the blade, $\Omega$ is the rotational speed of the rotor, $\bm{e}_x$ and $\bm{e}_\theta$ are the unit vectors in the axial flow and rotor rotating directions, respectively. In the present model, the axial and azimuthal components of the flow velocity, i.e. $u_x$ and $u_\theta$ are computed at the leading edge of the blade. Generally, the leading edge point LE does not coincide with any background nodes. In the present work, the smoothed discrete delta function (i.e. the smoothed four-point cosine function) proposed by \citet{yang2009smoothing} is employed to interpolate the flow velocity at the leading edge of the blade from the background grid nodes. The forces computed on each grid node of the leading edge are then uniformly distributed onto the corresponding actuator surface meshes with the same radius from the rotor center. The forces on the blade actuator surface are then distributed to the background grid nodes for the flow field using the same discrete delta function employed in the velocity interpolation process. The stall delay model developed by \citet{du19983} and the tip-loss correction proposed by \citet{shen2005tip} are employed to take into account the three-dimensional effects. \\
\indent In the actuator surface model for the nacelle, the nacelle geometry is represented by the actual surface of the nacelle with distributed forces. The force on the actuator surface is decomposed into two parts: the normal component and the tangential component. The normal component of the force is computed in a way to satisfy the non-penetration condition, which is similar to the direct forcing immersed boundary methods \citep{uhlmann2005immersed}, as follows:
\begin{equation} \label{eq:Fn_ASN}
F_n=\frac{\left(\bm{u}^d(\bm{X})-\bm{\tilde{u}}(\bm{X})\right)\cdot \bm{n}(\bm{X})}{\Delta t},
\end{equation}
where  $\bm{X}$ represents the coordinates of the nacelle surface mesh, $\bm{u}^d(\bm{X})$ is the desired velocity on the nacelle surface, $\bm{n}(\bm{X})$ is the unit vector in the normal direction of the nacelle, $\bm{\tilde{u}}$ is the velocity estimated from the previous flow field using an explicit Euler scheme, and $\Delta t$ is the time step. 
The tangential force is assumed to be proportional to the incoming velocity $U$ and is computed as follows:
\begin{equation}\label{eq:Ft_ASN}
F_{t}=\frac{1}{2}c_{f}U^2
\end{equation}
where $c_{f}$ is calculated from the empirical relation proposed by F. Schultz-Grunow ~\cite{schlichting2003boundary}.  The direction of the tangential force is determined by the local tangential velocity. 
The smoothed discrete delta function is employed for the velocity interpolation and force distribution as in the actuator surface model for blades. Validations of the proposed actuator surface models for turbine blades and nacelle can be found in \citep{yang2017actuator}. 
%
%
%
%
%
%
\section{Test cases and computational setup}\label{sec:cases}
\begin{figure}
  \begin{center} 
     \includegraphics[width=\textwidth]{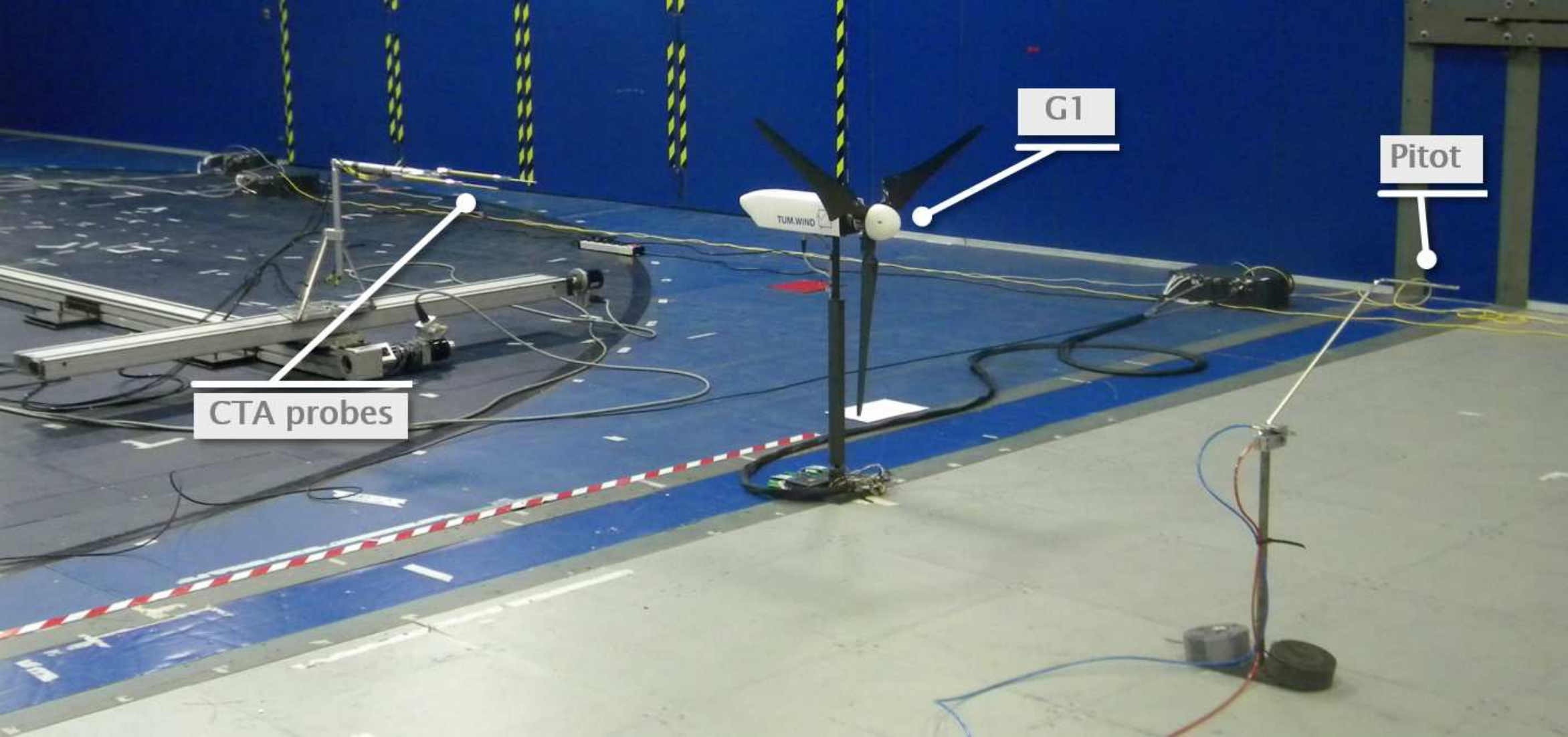}
     \caption{\label{fig:exp setup}Experimental setup in the wind tunnel} 
   \end{center}
\end{figure}

\begin{figure}
   \begin{center}
      \includegraphics[width=\textwidth]{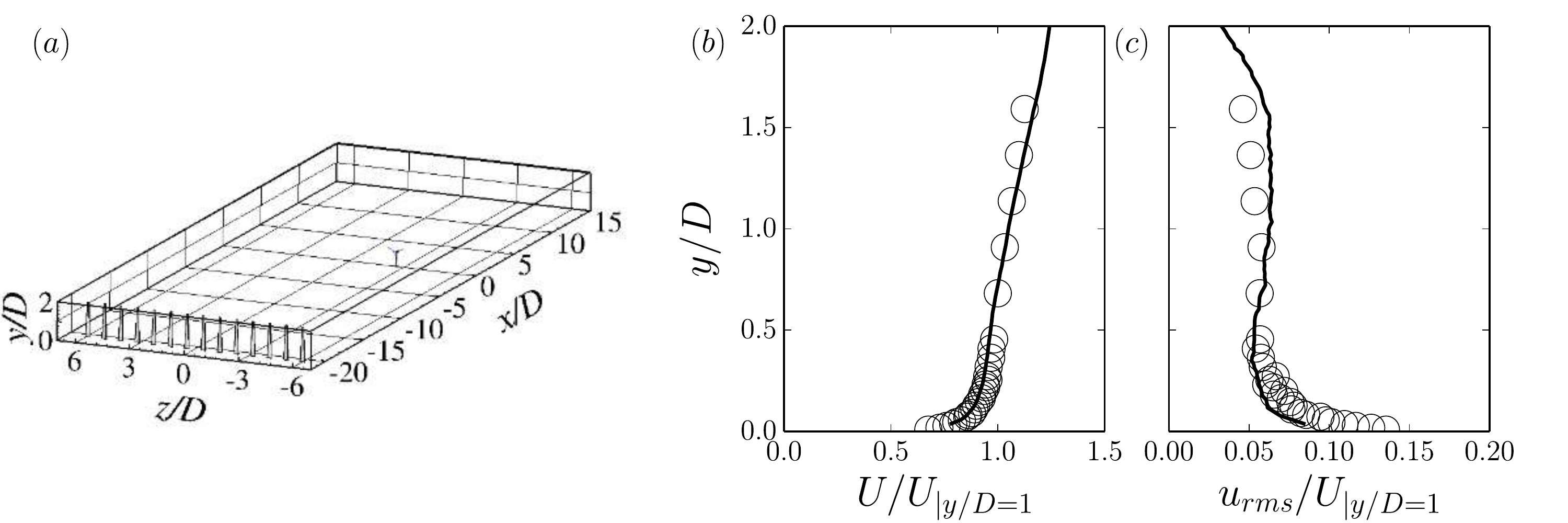}
      \caption{\label{fig:inflow}(a) Sketch of the wind tunnel domain with spires at the inlet $(x/D=-20.5)$ and turbine located at ($x/D=0$ and $z/D=-2.5$). Comparison of the vertical profiles at $x/D =3.5$ from a precursor simulation (solid lines) with measurements (circles) for an empty wind tunnel for (b) mean streamwise velocity and (c) rms streamwise velocity.}
   \end{center}
\end{figure}

\indent Here we discuss the computational details and overall setup for the two wind tunnel experiments where measurements of the velocity flow field are taken at two different operating conditions of Region 2 and Region 3 with the corresponding setup shown in Table \ref{tbl:cond}.  \\ 
\begin{table}
  \begin{center}
\def~{\hphantom{0}}
  \begin{tabular}{lccccc}
       Region   &  $\beta$ [deg] & Tip-speed ratio &  $C_p$  & $C_T$ & $Re_D = U_{hub}D/\nu $\\
       2       & 1.4  & 8.1 & 0.45 & 0.79 & $4.4\times 10^5$\\
       3      & 7.0  & 8.1 &  0.25 & 0.3 & $5.1 \times 10^5$\\  
  \end{tabular}
  \caption{Turbine operating condition parameters and performance. }
  \label{tbl:cond}
  \end{center}
\end{table}
\indent The wake behind the rotor was traversed and measured using 2 Tri-axial fiber-film probes \textit{Dantec 55R91}, while wind speed was also constantly measured by means of a Pitot tube located 1.5D in front of the rotor disk and at hub height, as shown in Fig.~\ref{fig:exp setup}.  The pressure signals were acquired by a \textit{MENSORCPT-6100 FS.=0.36PSI} Pitot transducer. \\
\indent Simulations are designed to reproduce the real wind tunnel environment based on the experiments including the spires at the inlet to induce a turbulent boundary layer that resembles an atmospheric boundary layer. 
A sketch of the entire wind tunnel domain with upstream spires and turbine is shown in Fig. \ref{fig:inflow}(a). A precursor LES of the entire wind tunnel without the turbine but resolving the details of the spire geometry using the CURVIB method is performed to acquire inflow conditions for subsequent turbine simulations \citep{foti2017use}. The dimensions of the wind tunnel domain are as follows: spanwise width $L_z/D=12.5$, height $L_y/D=3.5$, and a streamwise length $L_z/D = 31.8$. The computational domain is discretized with 600$\times$180$\times$1200 cells in spanwise, vertical and streamwise directions, respectively. Fourteen spires, resolved using with CURVIB method, with a height $H/D = 1.8$ and width $S/D=0.45$ are placed one meter apart symmetrically around the wind tunnel centerline.  The trailing edge of the spires is placed at a streamwise distance of $2.3$ diameters downstream of the domain inlet. Uniform inflow velocity $U_b$ is applied at the inlet.  A wall model for smooth walls \citep{yang2015large} is employed on the top, bottom and side walls of the domain, while boundary conditions on the spires are applied on the grid nodes in the vicinity of the spires using wall model reconstruction as described in \citet{kang2012numerical}.
 Further analysis of the precursor simulation and the generated inflow resemblance to an atmospheric boundary layer can be found in \citet{foti2017use}. The instantaneous flow field is acquired and saved at the location upstream of the turbine at $x/D = -1.5$, and is fed into the wind turbine simulations.   \\
\indent The mean streamwise velocity and rms (root-mean-square) streamwise velocity from the precursor simulation with the spires without the wind turbine at $x/D=3.5$ are shown in Fig. \ref{fig:inflow}(b) and Fig. \ref{fig:inflow}(c), respectively, where both the simulations and experimental measurements converge to the same profile. At this location the velocity at turbine hub height is $U_{hub}/U_b=1.0$, the incoming shear velocity is $u_\tau/U_b = 0.03$, and boundary layer height is $\delta/D = 1.8$, the same as the height of the spires. The Reynolds number based on $U_{hub}$ and $D$ is $4.0\times 10^5$. The value of Reynolds number of the empty wind tunnel simulations and turbine simulations in both operating conditions are beyond the value for which Reynolds number independence for turbine wakes is observed \citep{chamorro2012reynolds}.  The so-generated inflow flow fields from this precursor simulation are employed to provide inflow conditions in the simulations for both turbine operating conditions without running additional precursor simulations to match the specific values of Reynolds number. \\
\indent The turbine simulations are performed in a domain $12.5D\times 2.7D\times 16D$ in spanwise, vertical and streamwise directions, respectively.  The numbers of grid nodes are 272$\times$120$\times$589 in spanwise, vertical and streamwise directions, respectively, resulting in a grid spacing $D/50$ in all three directions.  This resolution is comparable to other studies using actuator-type models \citep{ivanell2010stability, troldborg2010numerical} and is sufficient to capture the main characteristics of the turbine wake as shown in Fig. \ref{fig:exp_6ms} and Fig. \ref{fig:exp_7ms} where we compare the computed results with experimental measurements. 
With this domain, the turbine is placed at the same location $x_0/D=20.5$ downstream of the spires and $z_0/D=-2.5$ as prescribed in the experiments. Similar to the precursor simulations, top, bottom and side walls of the domain utilize a wall model for smooth walls. 
Two simulations for each turbine operating conditions based on the experimental setup shown in Table \ref{tbl:cond} are carried out: one with the actuator surface models for blades and nacelle (R2-BN and R3-BN for Region 2 and Region 3, respectively); the other one with actuator surface model for blade only (R2-B and R3-B for Regions 2 and 3, respectively).
%
%
%
%
%
%
\section{Results}\label{sec:results}
\indent We present and analyze the computational and experimental results in this section. In section \ref{sec:Time-averaged}, we first compare the computed profiles with measurements at different downstream locations and show the time-averaged flow field and turbulence statistics for simulations with and without a nacelle model for the two different turbine operating conditions. We then present instantaneous flow fields for different cases and analyze the power spectral density in section \ref{sec:instantaneous}, and investigate the wake meandering using a filtering technique and dynamic mode decomposition method in section \ref{sec:meander} and section \ref{sec:dmd}, respectively.
%
%
\subsection{Time-averaged characteristics}\label{sec:Time-averaged}
\begin{figure}
   \begin{center}
       \includegraphics[width=\textwidth]{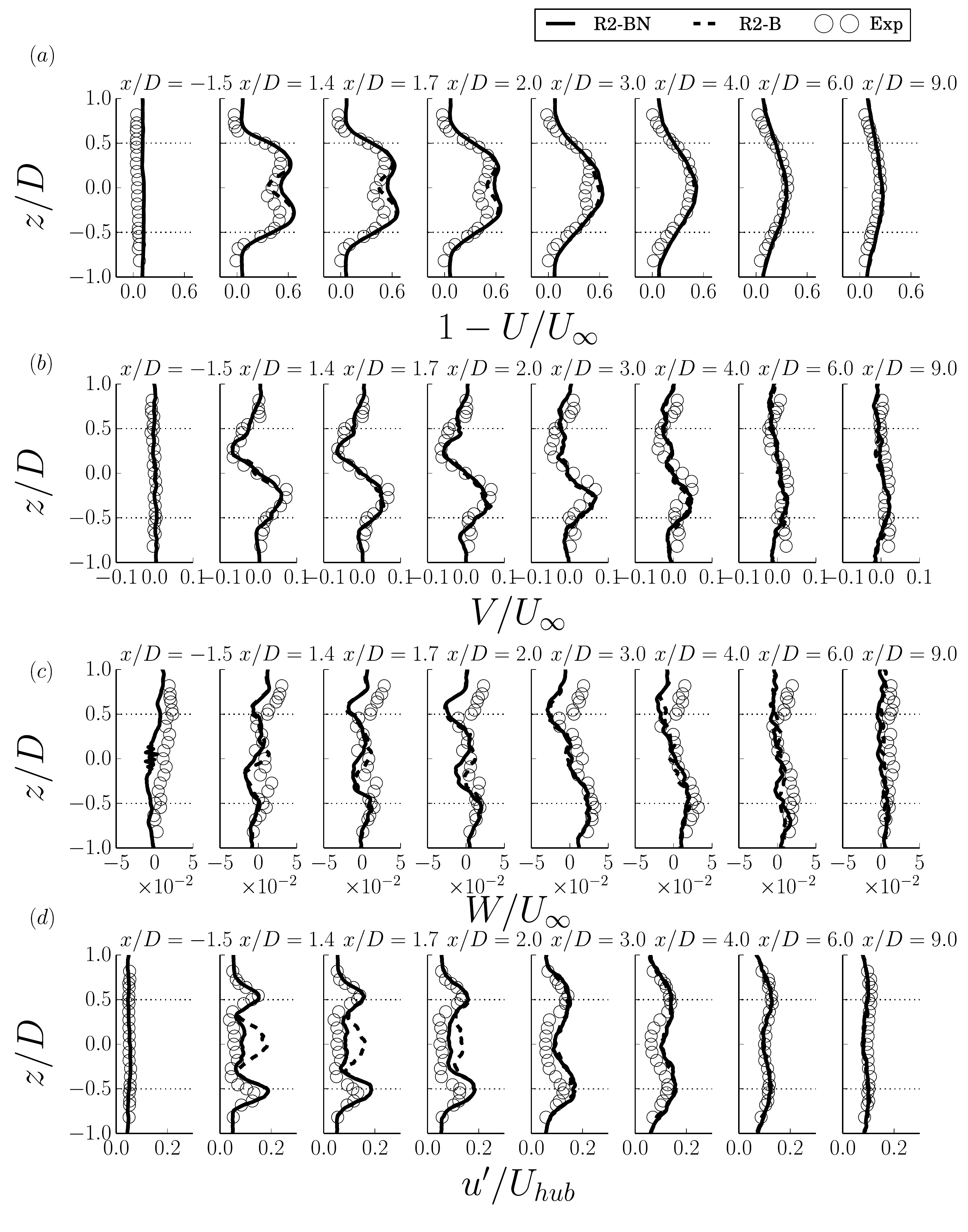}
       \caption{\label{fig:exp_6ms} Comparisons of the computed and measured spanwise profiles for Region 2 (a) streamwise velocity deficit, (b) vertical velocity, (c) spanwise velocity, and (d) streamwise RMS velocity $u^\prime$.  The horizontal dotted lines at $z/D = \pm 0.5$ are the tip positions. }
   \end{center}
\end{figure}

\begin{figure}
   \begin{center}
       \includegraphics[width=\textwidth]{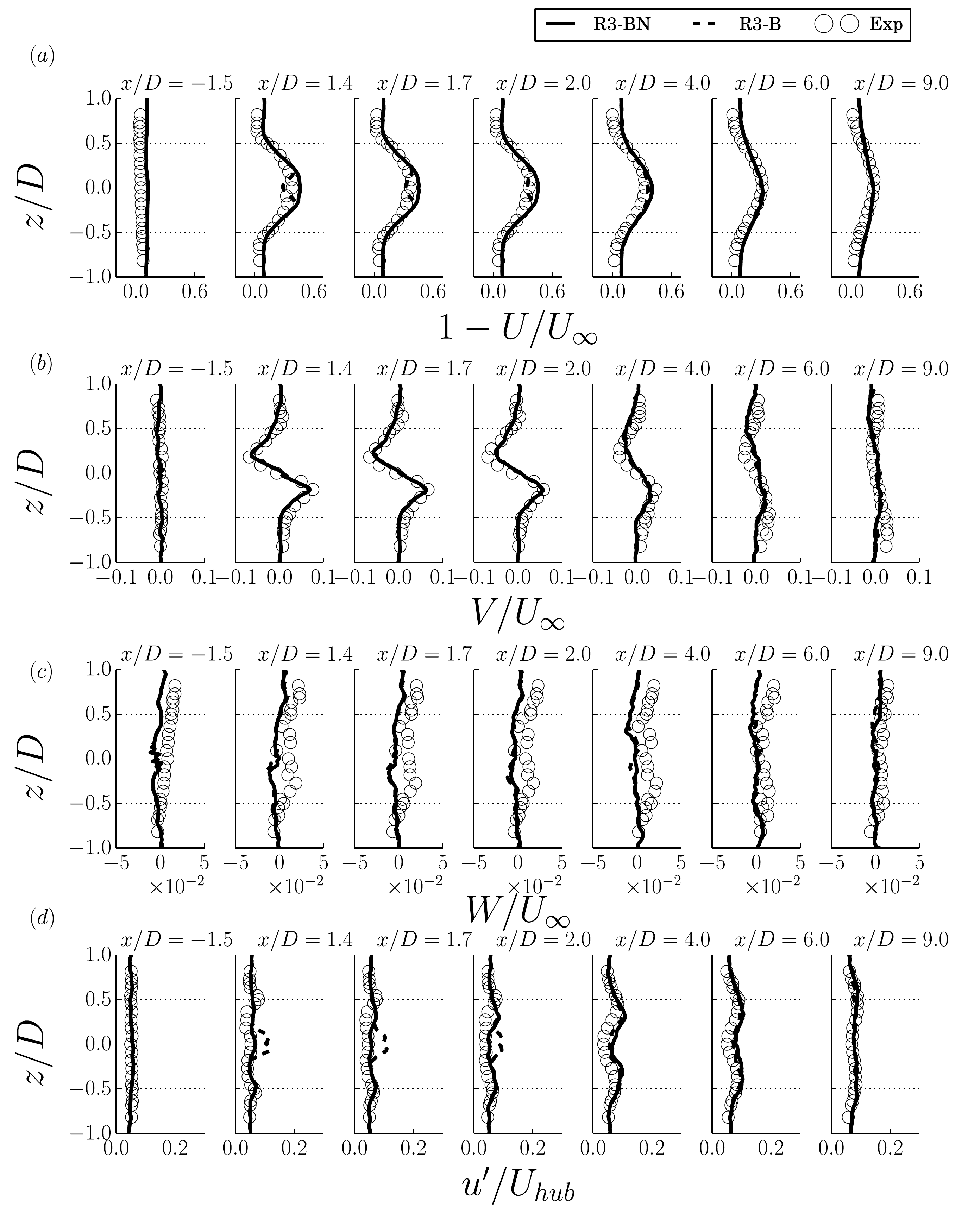}
       \caption{\label{fig:exp_7ms} Comparisons of the computed and measured spanwise profiles for Region 3 (a) streamwise velocity deficit, (b) vertical velocity, (c) spanwise velocity, and (d) streamwise RMS velocity $u^\prime$.  The horizontal dotted lines at $z/D = \pm 0.5$ are the tip positions.}
   \end{center}
\end{figure}

\begin{figure}
   \begin{center}
      \includegraphics[width=\textwidth]{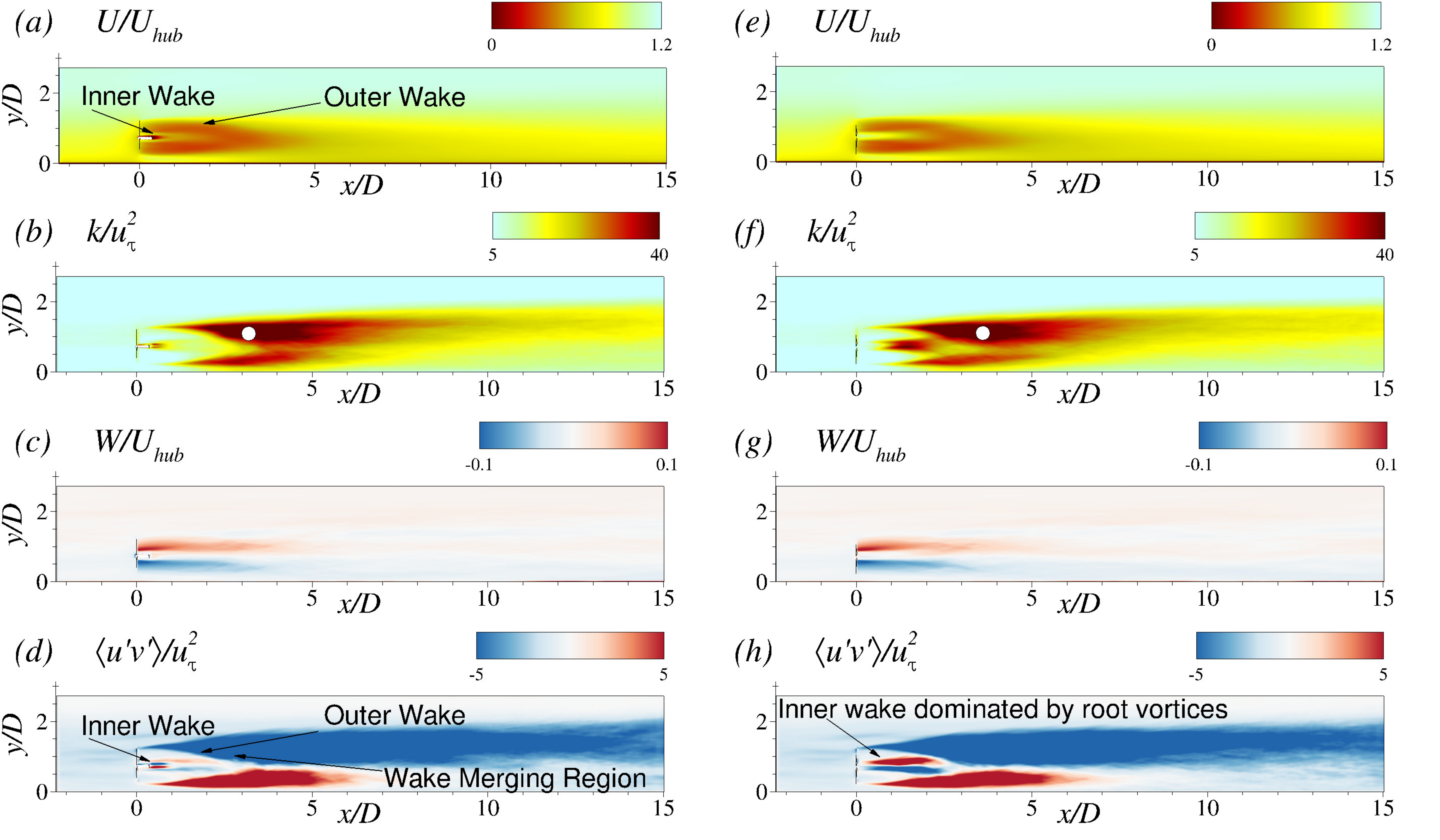}
      \caption{\label{fig:6_avg} Comparisons of vertical centerplane contours between simulations with (left images) and without (right images) a nacelle model for turbine operating in Region 2 for (a) and (e) time-averaged downwind velocity, (b) and (f) turbulence kinetic energy, (c) and (g) time-averaged spanwise (rotational) velocity, and (d) and (h) downwind-vertical Reynolds shear stress. Notations demarcating the inner and outer wake are shown on Fig. (a), (d), and (h).}
   \end{center}
\end{figure}

\begin{figure}
   \begin{center}
      \includegraphics[width=\textwidth]{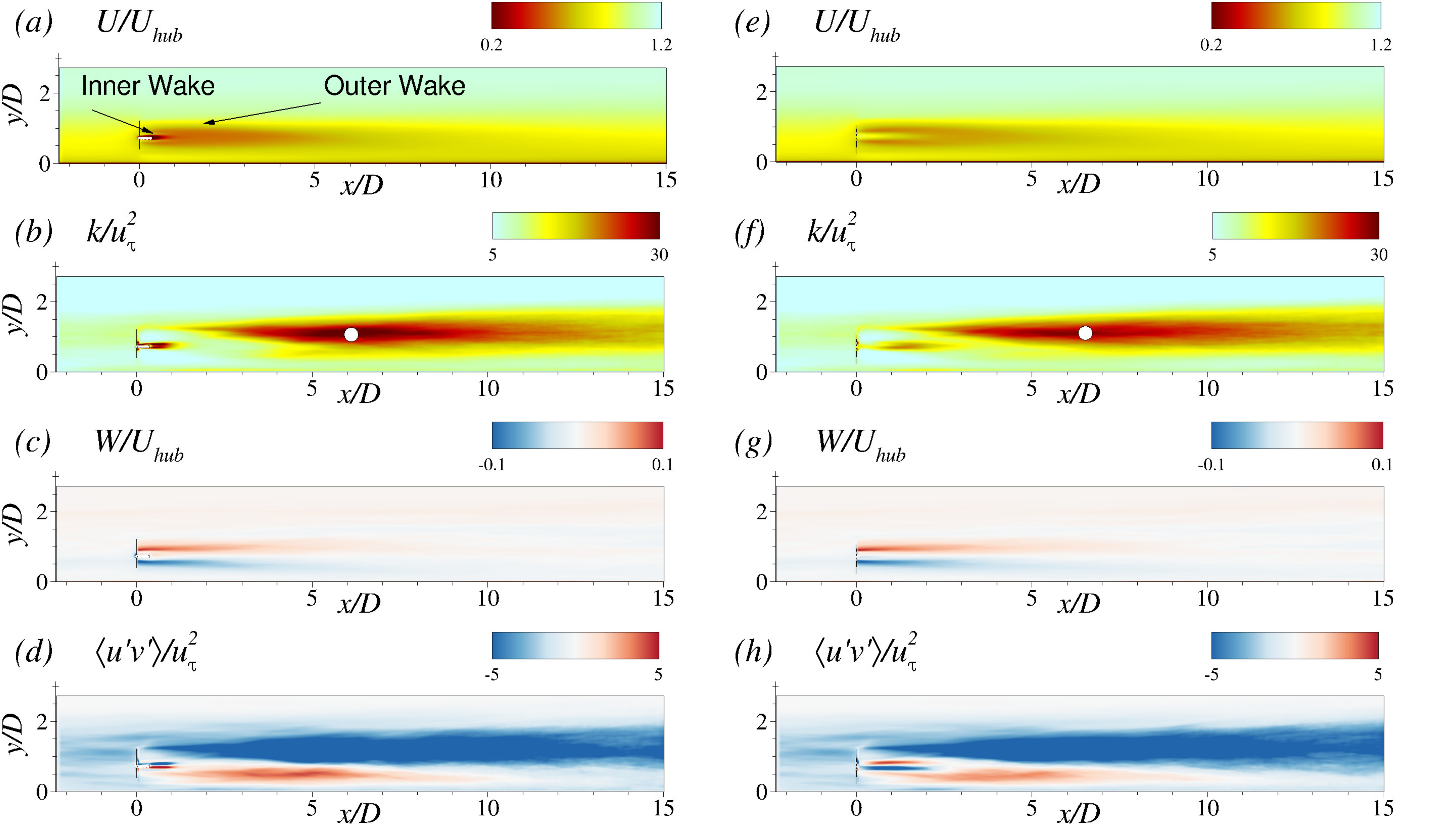}
      \caption{\label{fig:7_avg} Comparisons of vertical centerplane contours between simulations with (left images) and without (right images) a nacelle model for turbine operating in Region 3 for (a) and (e) time-averaged downwind velocity, (b) and (f) turbulence kinetic energy, (c) and (g) time-averaged spanwise (rotational) velocity, and (d) and (h) downwind-vertical Reynolds shear stress. Notations demarcating the inner and outer wake are shown on Fig. (a). }
   \end{center}
\end{figure}

\begin{figure}
   \begin{center}
      \includegraphics[width=\textwidth]{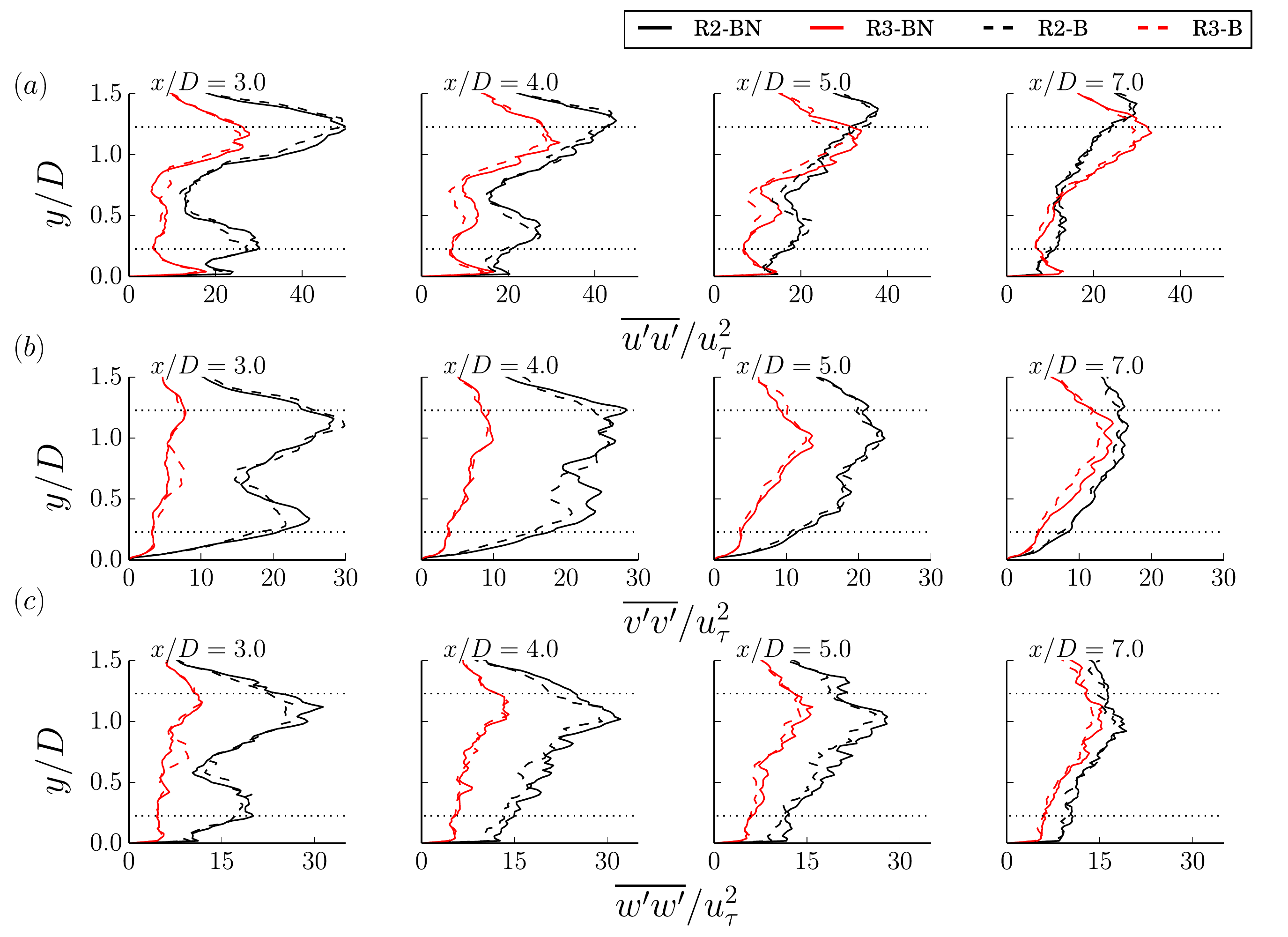}
      \caption{\label{fig:turb_uu} Vertical profiles of the turbulence variances for (a) $\overline{u^\prime u^\prime}/u_\tau^2$, (b) $\overline{v^\prime v^\prime}/u_\tau^2$, and (c) $\overline{w^\prime w^\prime}/u_\tau^2$ non-dimensionalized $u_\tau^2$, where $u_{\tau}$ is the friction velocity of the incoming turbulent boundary layer flow.  The horizontal dotted line show the bottom and top tip position at $y/D=0.23$ and $1.23$, respectively.}
   \end{center}
\end{figure}

\indent \indent To evaluate the capability of the actuator surface models in predicting the wake from the model wind turbine, we compare the computational results with the experimental measurements for the turbine operating in Region 2 and Region 3 in Fig. \ref{fig:exp_6ms} and Fig. \ref{fig:exp_7ms}, respectively.
Figure \ref{fig:exp_6ms}(a) shows the comparisons of the spanwise profiles of the mean streamwise velocity deficit, $1-U/U_\infty$ across the wake of the turbine.  In the profiles closest to the turbine at $x/D=1.4$ and $x/D=1.7$, the effect of the nacelle manifests itself as an increase of the streamwise velocity deficit along the centerline.  The employed nacelle model captures this inner wake feature from the nacelle, but the absence of a nacelle model results in an unphysical jet at the center of the profiles. On the other hand, the shear layer of the outer wake, mainly caused by the forces on the blades extending to $z/D \pm 0.5$ (along the horizontal dotted lines in Fig. \ref{fig:exp_6ms}) is captured well by both simulations with and without a nacelle model.  
Further downstream of the rotor, the inner wake of the nacelle begins to merge with the outer wake and the streamwise velocity deficit begins to flatten along the centerline.  The unphysical jet in the simulation without the nacelle model, created by the absence of the nacelle model, begins to dissipate at further downstream locations as the corresponding agreement with experimental measurements becomes better.   Far from the rotor $x/D > 4.0$, the wake recovery becomes slow for all the cases.
Figure \ref{fig:exp_6ms}(b) shows the comparisons of the mean vertical velocity $V$ which represents the rotational velocity on this plane.  Both simulations with and without a nacelle model agree well with the experimental measurements.  Close to the turbine, the rotation of the blades imparts a strong mean rotation on the inner wake that peaks just away from the center. Despite the high rotational velocity occurring near the nacelle boundary, the nacelle model does not affect the mean vertical velocity, which is due only to the rotation of the rotor.  Farther downstream the mean rotation quickly dissipates.  
The mean spanwise velocity $W$ is shown in Fig. \ref{fig:exp_6ms}(c).  The simulations compare well with the measurements despite the small magnitude of this velocity component.  
The final experimental measurement for this operating regime, shown in Fig. \ref{fig:exp_6ms}(d),  is the streamwise rms (root-mean-square) velocity $u^\prime$.  
The cases with and without a nacelle model capture the magnitude of the turbulence intensity of nearly $0.2$ within the tip shear layer showing agreement with the experiments. Near the centerline and $x/D=1.4$, both the measurements and the simulation with a nacelle model capture the shear layer of the inner wake with a slight increase of the turbulence intensity. The simulation without the nacelle model shows an unphysical increase peaking at the centerline.   As the wake progresses downstream, the centerline peak of turbulence intensity in the simulation without a nacelle model decreases and converges to the measurement profiles at $x/D=3$. Furthermore, as the inner and outer wake shear layer expand, the turbulence intensity along the tip positions relaxes to about $u^\prime/U_{hub} = 0.1$ where the measurements, simulations with and without a nacelle model agree well with each other.\\
\indent The spanwise profiles of the mean velocity deficit of the measurements and simulations for the turbine operating in Region 3 are shown in Fig. \ref{fig:exp_7ms}(a).  In comparison with the turbine operating in Region 2, significantly less velocity deficit is in the region near blade tips at near wake locations ($x/D=1.4, 1.7, 2$). Near the centerline the velocity deficit is lower and the inner wake of the nacelle is smaller with a flattened profile. Similarly, the simulation without a nacelle model under predicts the velocity along the centerline until $x/D > 2$.  Far away from the turbine rotor, the wake relaxes and all of the simulation cases are able to accurately simulate the wake.  
The comparison of the vertical velocity $V$ profiles (rotational component on this plane) is shown in Fig. \ref{fig:exp_7ms}(b). In comparison with the turbine operating in Region 2, the magnitude of rotational velocity near the blade tips are smaller for the turbine operating in Region 3 at near wake locations ($x/D=1.4, 1.7, 2$). The spanwise velocity $W$ profiles shown in Fig. \ref{fig:exp_7ms}(c), on the other hand, are similar to the profiles for the turbine operating in Region 2 and both simulations with and without nacelle model show good agreement with the measurements.      
Finally, the streamwise rms velocity is shown in Fig. \ref{fig:exp_7ms}(d). Similar to what we observed in Fig. \ref{fig:exp_6ms}(d) for the turbine operating in Region 2, the lack of a nacelle model introduce peaks on the streamwise turbulence intensity in the region near the rotor centerline at near wake locations ($x/D=1.4, 1.7, 2$). The tip shear layer is accurately simulated by all the simulations.   As the wake progresses downstream, differences in the mean flow field are not obvious.   \\     
\indent Next, we will investigate the spatial distribution of the mean flow field via contour plots. Mean flow and turbulence quantities for Region 2 simulations with and without a nacelle model on the vertical-streamwise centerline plane are shown in Figs. \ref{fig:6_avg}(a)-(d) and Figs. \ref{fig:6_avg}(e)-(h), respectively.
The mean streamwise velocity $U/U_{hub}$ for the simulation with a nacelle model in Fig. \ref{fig:6_avg}(a), show the outer wake formed by the turbine rotor and the inner wake behind the nacelle (outer and inner wake are demarcated on figure). In the mean velocity contour shown in Fig. \ref{fig:6_avg}(e) for the simulation without a nacelle model, a jet exists along the centerline, as discussed in the spanwise profile shown in Fig. \ref{fig:exp_6ms} and Fig. \ref{fig:exp_7ms} and seen in simulations in \citet{kang2014onset}, exists as a consequence of not having the blockage effect of a nacelle body. The outer wake shear layer at the top tip position immediately downstream of the turbine is very sharp and slowly expands.  The inner wake forms a core region where the hub vortex begins to grow. \\
\indent The turbulence kinetic energy $k/u_\tau^2$ (TKE) contours from the simulations with and without a nacelle model are shown in Fig. \ref{fig:6_avg}(b) and Fig. \ref{fig:6_avg}(f), respectively.  
As seen, the intense turbulence regions start at the rotor tips. From the TKE of the simulation with a nacelle model, the outer wake and the expanding inner wake are observed.  The inner wake expands outwards towards and intercepts the outer wake a few diameters downstream of the turbine.  Near this intersection, there is a marked increase in TKE caused by the interaction of the expanding inner wake caused by the hub vortex and the outer wake.  The growth of a large region of high TKE in the tip shear commences near the intersection of the inner wake shear layer.    
In \citet{kang2014onset} and with further evidence in \citet{foti2016wake}, this region of high TKE is associated with the onset of wake meandering.
For the simulation without a nacelle model, a hub vortex is still formed along the centerline of the turbine but no nacelle inner wake is formed.  The increase in streamwise velocity causes the hub vortex to remain strong and columnar.  Expansion of the hub vortex towards the remnants of the tip vortices at the tip position does not occur.  Instead, the onset of wake meandering occurs with less turbulence energy and further downstream. The location of the maximum TKE with $k/u_{\tau}^2=52.9$ occurs at $x/D=3$ for the simulation with a nacelle model; while it is found to be at $x/D=3.5$ with $k/u_{\tau}^2=52.8$ for the simulation without a nacelle model. The locations are shown on the TKE contours in Fig. \ref{fig:6_avg}(b) and Fig. \ref{fig:6_avg}(f) as large circles.\\  
\indent The mean rotational velocity in the wake is expressed as the mean spanwise velocity $W/U_{hub}$ where positive is the direction into the page.  For reference, the turbine rotates clockwise, thus the top blade tip is moving out of the page.  In Fig. \ref{fig:6_avg}(c) and Fig. \ref{fig:6_avg}(g), showing the simulations with and without a nacelle model, respectively, a mean rotational rate is present at turbine near wake locations. Both figures show a wake that is rotating counter-clockwise (opposite the rotation of the turbine blades) with the strongest rotation occurring near the root of the blade.  By the tip of the blade, the rotation is nearly zero.  By $x/D=5$, the mean rotation in the wake has dissipated to be negligible. \\
\indent The streamwise-vertical Reynolds stress $\overline{u^\prime v^\prime}/u_\tau^2$ for simulations with and without a nacelle model, shown in Fig. \ref{fig:6_avg}(d) and Fig. \ref{fig:6_avg}(h), respectively, shows the turbulence shear stress and mixing that occurs in the shear layers of the inner and outer wake.  There are several regions of intense shear stress in the simulation with a nacelle model:
i.) The turbine tip shear layer which extends into the wake as a thin layer for the first few diameters downstream where the tip vortices remain coherent occupying the interface between the fast moving outer flow and the turbine wake.  As the tip vortices start to breakdown the region of mixing begins to expand through entrainment of fluid from outside the wake;  
ii.) Behind the nacelle along the shear layer of the inner wake.  This region of high Reynolds stress promotes mixing of the slowest moving inner wake with the outer wake behind the turbine blades.  Moreover, this allows for entrainment of fluid into the inner wake and rapid expansion of the inner wake and hub vortex.  Encompassing the inner wake is a region of opposite-sign low Reynolds shear stress that originates from the root of the turbine blades;  
iii.) Much like the TKE, there is a large region of high Reynolds shear stress that appears downstream along the tip positions and down towards the center of the wake. Here, the inner and outer wake merge together and intense mixing occurs.  As the wake meandering region begins at approximately $x/D=2$ downstream, the sign of the Reynolds stress indicates that slower moving fluid in the wake will mix and entrain fluid from the outer wake to initiate wake recovery. 
The streamwise-vertical Reynolds stress of the simulation without the nacelle model (Fig. \ref{fig:6_avg}(h))  provides more evidence of the strong columnar hub vortex which is dominated by strong Reynolds shear stress originating from the root of the turbine blades.  While the Reynolds shear stress is relatively high in the inner wake, it is dominated by the root vortices which does not promote a similar laterally expanding inner wake from a nacelle as in the simulation with the nacelle model.  The Reynolds shear stress from the root vortices indicates that mixing of the faster moving fluid from behind the turbine blades towards the centerline does not occur as effectively as in the simulation with a nacelle model.  The centerline jet creates a non-physical acceleration of the streamwise flow promoting a stable columnar vortex.  Further downstream, the increased Reynolds stress behaves similar to the nacelle model cases but with less turbulence mixing. Overall Reynolds shear stress comparisons of the simulations with and without nacelle suggest quantitative differences in the intensity of the far wake meandering region.  This important issues will be discussed extensively in subsequent sections of this paper. \\
\indent Figures \ref{fig:7_avg}(a)-(d) and Figs. \ref{fig:7_avg}(e)-(h) show contours of mean flow quantities for the turbine operating in Region 3 on the vertical-streamwise centerplane for simulations with and without a nacelle model, respectively.  The mean streamwise velocity contours in Fig. \ref{fig:7_avg}(a) shows similar outer and inner wakes patterns as that in Fig. \ref{fig:6_avg}(a) for the turbine operating in Region 2.
However, in comparison with that for the turbine operating in Region 2, the strength of the velocity deficit in the outer wake for the turbine operating in Region 3 is significantly less because of smaller power and thrust coefficients that resulted from the large pitch angle of the blade. The reduced velocity deficit enables faster recovery of the wake with less expansion in the radial direction for the Region 3 case. Similar to the Region 2 cases, lack of a nacelle model also results in an unphysical centerline jet which affects the wake development. \\ 
\indent The TKE $k/u_\tau^2$ contours are shown in Fig. \ref{fig:7_avg}(b) and Fig.  \ref{fig:7_avg}(f), for simulations with and without a nacelle model, respectively.  As seen, the maximum TKE in the top tip shear layer is higher and the region with high TKE is wider for the simulation with a nacelle model in comparison with that without a nacelle model.  The location of maximum TKE occurs at $x/D=5$ downstream for the simulation with a nacelle model, while the corresponding location for the simulation without a nacelle model occurs further downstream at $x/D=5.5$.  
Comparing the TKE contours in Fig. \ref{fig:7_avg}(b) and Fig. \ref{fig:6_avg}(b) for the two operating conditions, we see differences in both the magnitude of the TKE and the downstream location where the highest TKE occurs.  The region with high TKE for the turbine operating in Region 3 is longer and extends at further downstream locations but with significantly less maximum TKE in comparison with that in Region 2.
The rotation of the wake for the turbine operating in Region 3, presented in Fig. \ref{fig:7_avg}(c) and Fig. \ref{fig:7_avg}(g), respectively, is more confined and with a comparably weaker maximum rotation. However, the mean rotation in the wake takes about the same distance or even somewhat longer distance from the rotor to dissipate. The streamwise-vertical Reynolds stress $\overline{u^\prime v^\prime}/u_\tau^2$ is consistent with the results from the turbine operating in Region 2, but with lower turbulence mixing levels due to less velocity deficit and weaker wake meandering as will be shown in section~\ref{sec:meander}.  \\
\indent Turbulence intensities at far wake locations can be considered as footprints of wake meandering. In order to further evaluate the nacelle effects at far wake locations, we show in Fig. \ref{fig:turb_uu} different components of turbulence intensities at various downstream locations. For the Region 2 cases, the streamwise turbulence intensity computed from the case with a nacelle model is slightly higher than that without a nacelle model at $x/D=3, 4$ and $5$ locations. However, the differences are more significant for the other two components at $x/D=3$ and 4 locations. For the Region 3 cases, the streamwise turbulence intensity computed from the simulation with a nacelle model is larger than that without a nacelle model especially at locations $x/D=4$ and 5 from the hub height to the top tip position. The differences for the other two components of turbulence intensity, on the other hand, are very minor downstream. However, at $x/D=3$ near the hub height differences in radial and azimuthal turbulence intensities exist due to the formation of the unphysical hub vortex developed without a nacelle model. Comparing the turbulence intensities between the two different turbine operating conditions, we can see that at $x/D=3$ and 4 locations, the turbulence intensities are larger for the Region 2 condition for all the three components;  at $x/D=5$ and 7 locations; on the other hand, the streamwise turbulence intensity near the top tip location is larger for the Region 3 condition. Overall, Fig. \ref{fig:turb_uu} quantitatively indicates the significant effects of turbine operating condition and nacelle on turbulence intensity at far wake locations for this model wind turbine. \\
%
%
%
\subsection{Instantaneous flow fields and spectral analysis}\label{sec:instantaneous}
\begin{figure}
   \begin{center}
      \includegraphics[width=\textwidth]{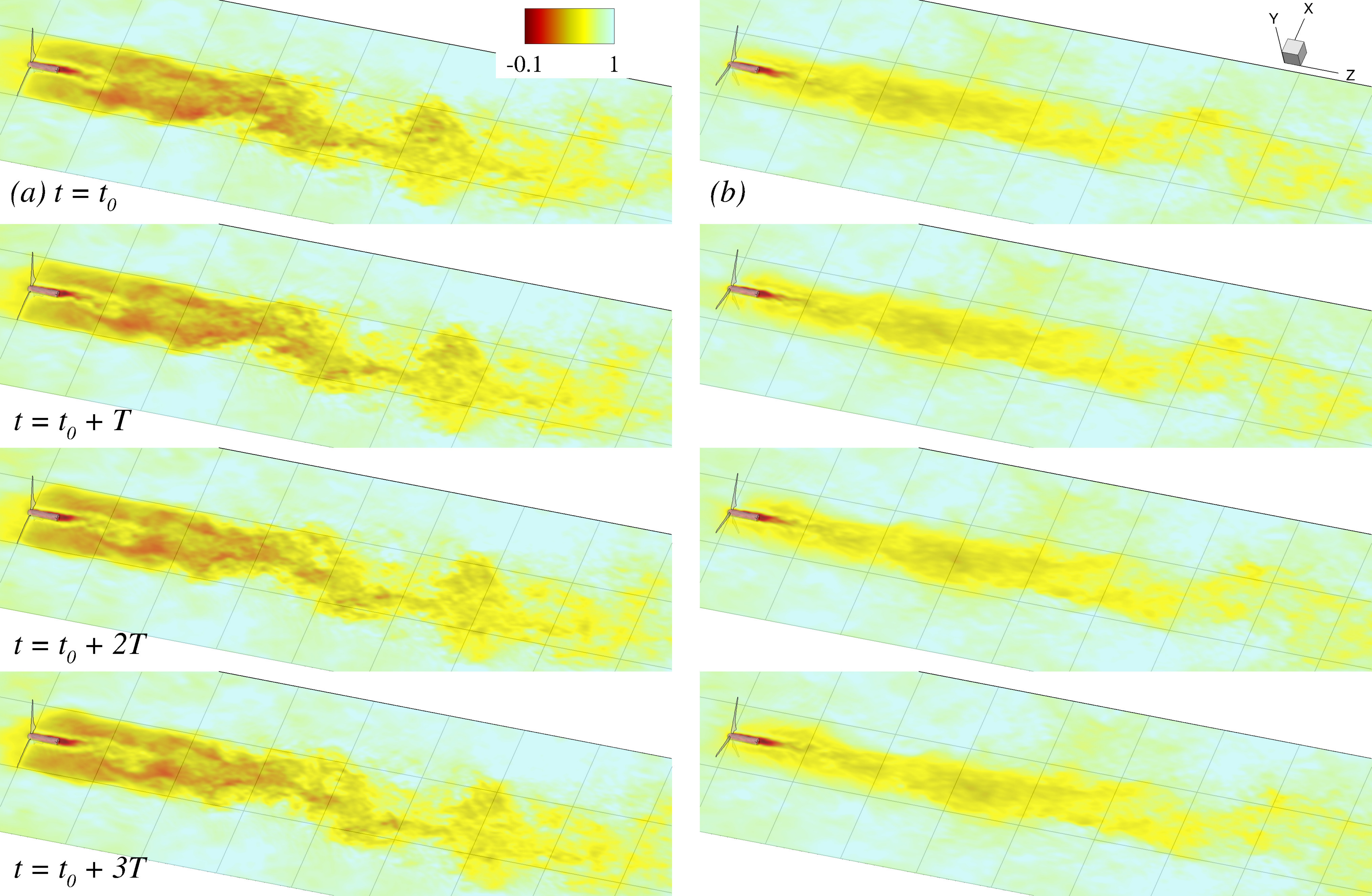}
      \caption{\label{fig:asn_inst} Contours of the instantaneous streamwise velocity at four successive time instances for (a) turbine operating in Region 2 (R2-BN) and (b) turbine operating in Region 3 (R3-BN).}
   \end{center}
\end{figure}

\begin{figure}
   \begin{center}
      \includegraphics[width=\textwidth]{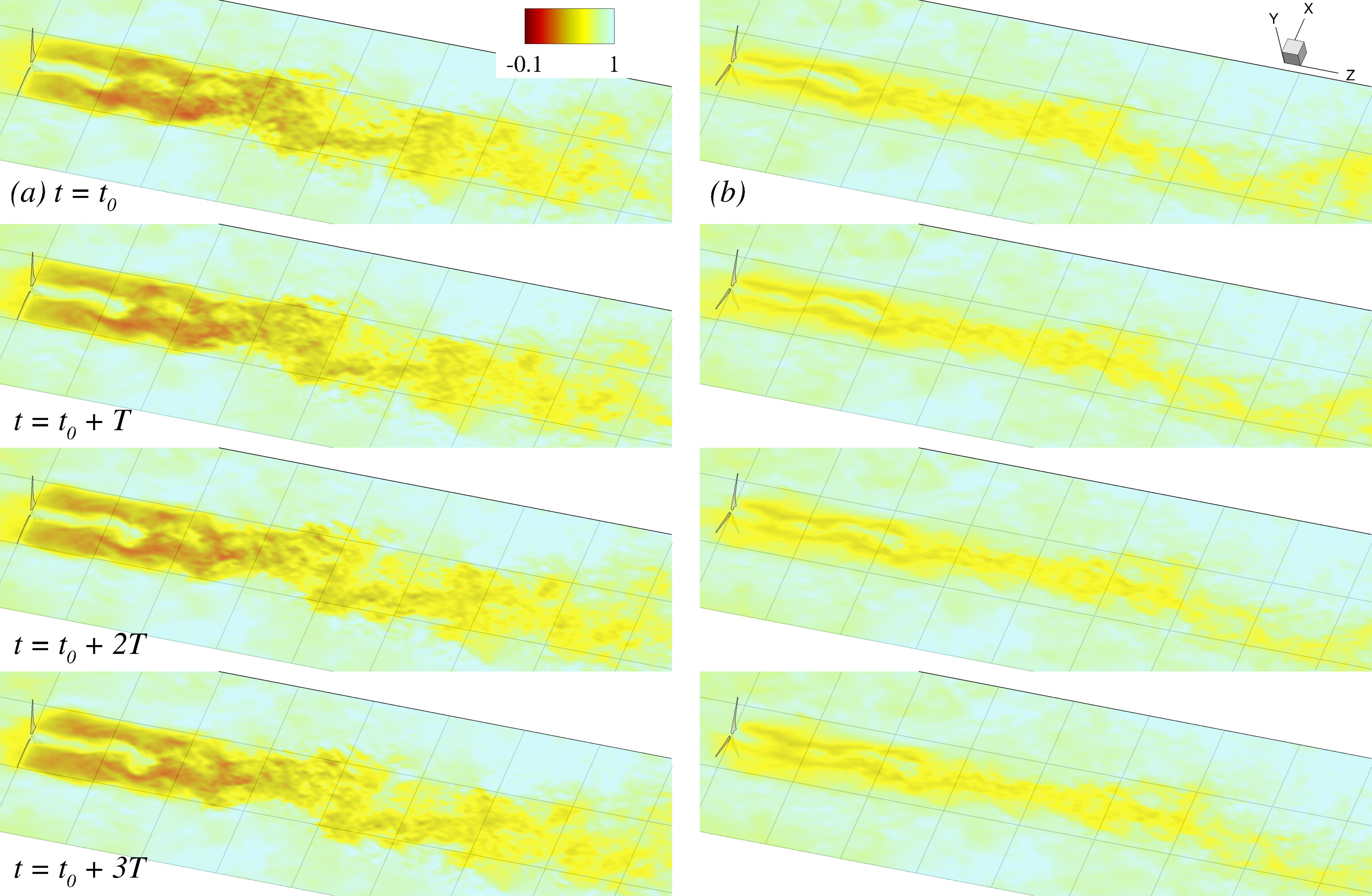}
      \caption{\label{fig:as_inst} Contours of the instantaneous streamwise velocity without a nacelle model at four instantaneous time instances for (a) turbine operating in Region 2 (R2-B) and (b) turbine operating in Region 3 (R3-B).}
   \end{center}
\end{figure}

\begin{figure}
   \begin{center}
      \includegraphics[width=\textwidth]{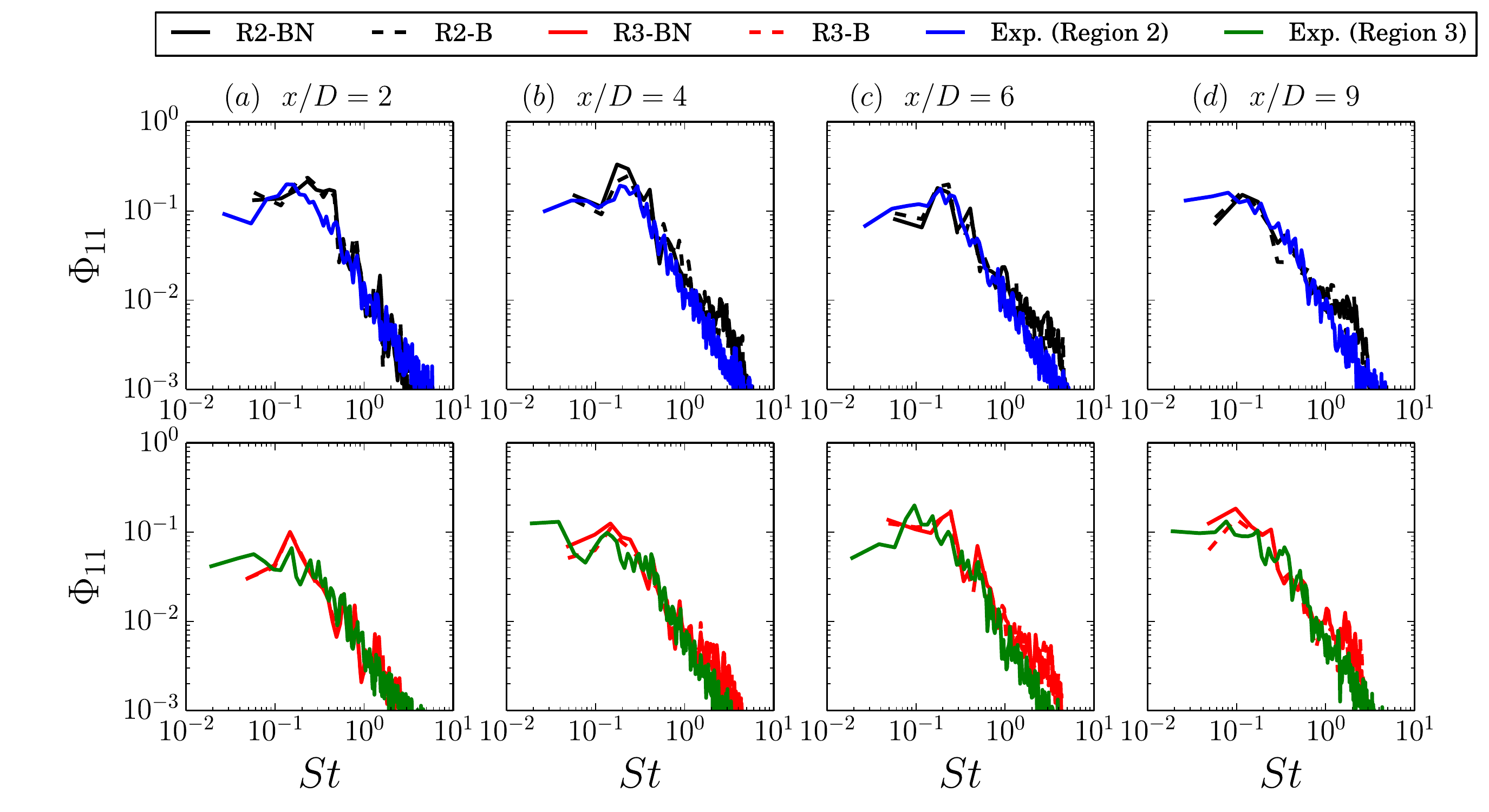}
      \caption{\label{fig:exp_psd} Comparisons of the Fourier power spectral density of the streamwise component of the velocity fluctuations $\Phi_{11}$ with that from measurements, with the top images and bottom images in Region 2 and Region 3, respectively, along the tip position $y/D = 0.72$, and $z/D=1$  (a) at $x/D=2$, (b) at $x/D=4$, (c) at $x/D=6$, and (d) at $x/D=9$.} 
   \end{center}
\end{figure}

\begin{figure}
   \begin{center}
      \includegraphics[width=\textwidth]{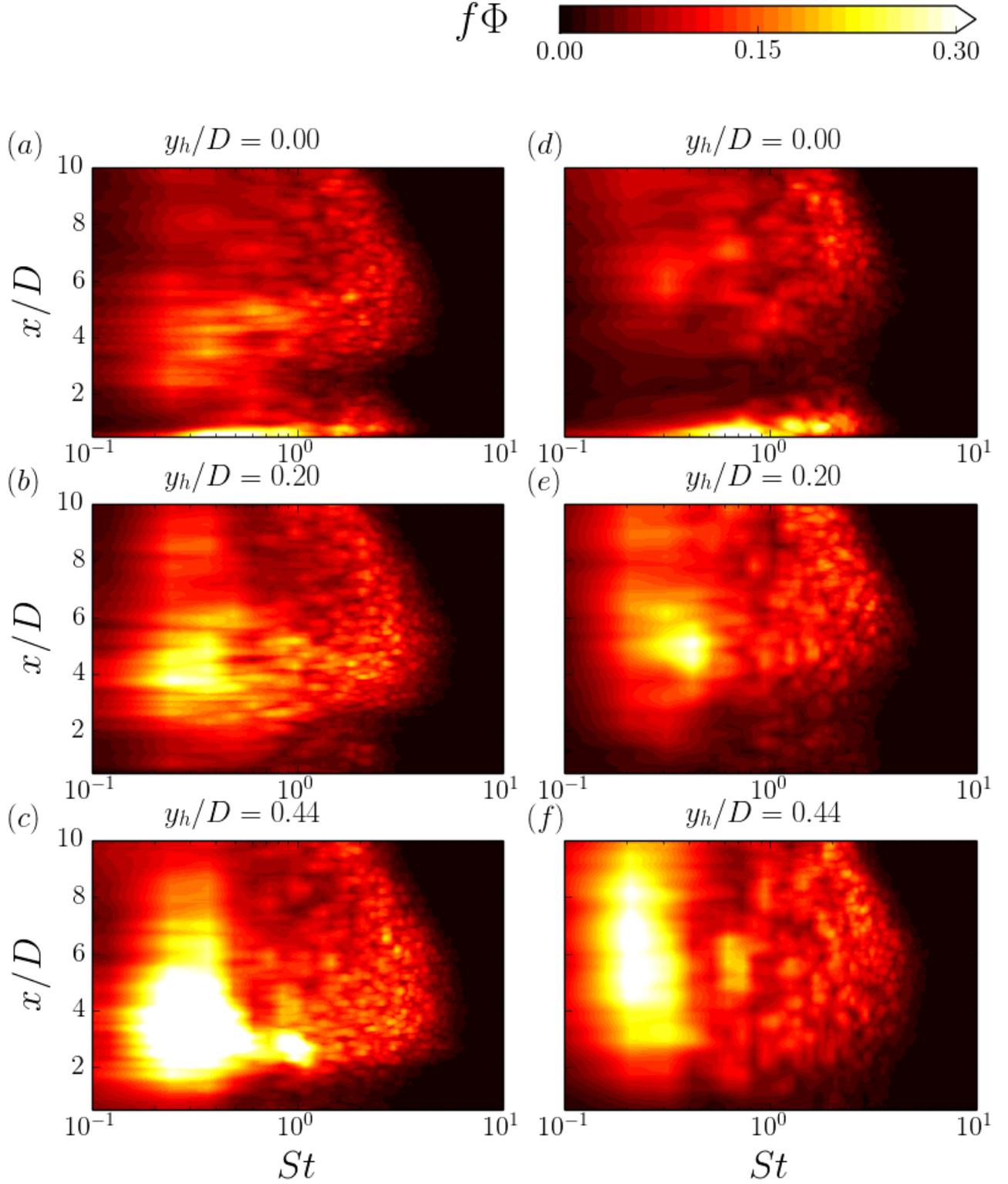}
      \caption{\label{fig:linefft} Contours of the Fourier power spectral density of the streamwise component of the velocity fluctuations pre-multiplied by frequency $f\Phi$ as a function of the axial direction, $x/D$ at several radial positions (a),(d) $y_h/D=0$, (b),(e) $y_h/D=0.2$ (c),(f) $y_h/D=0.44$. Left column (a)-(c): R2-BN and right column (d)-(f): R3-BN. } 
   \end{center}
\end{figure}

\indent In this section, we focus on the instantaneous flow fields to get an intuitive cognition on the different wake meandering patterns for different operating conditions and simulations with and without a nacelle model. The analysis of the mean flow field demonstrated that both the inner wake and outer wake form behind the turbine. In the mean sense, the inner wake expands into the outer wake and the turbulence kinetic energy increases substantially downstream.  
To further investigate the instantaneous evolution of the wake and substantiate the narrative presented above, several time instances of both operating conditions from the simulations with a nacelle are shown in Fig. \ref{fig:asn_inst}.  Each successive instance is separated by one rotor rotation period $T = 0.07$ s.  The instantaneous flow field from the Region 2 case, in Fig. \ref{fig:asn_inst}(a), contains an inner wake directly behind the nacelle and a large outer wake extending from the extent of the rotor blades.  Over the successive figures the slow precession of the hub vortex and the expansion of the inner wake becomes evident. Around $x/D = 2$, the inner wake begins to interact with the outer wake, and large lateral excursions of the outer wake mark the onset of the meandering.  By $x/D=4$ the full extent of the wake meandering has commenced.  After the emergence of the meandering, progressively larger spanwise displacements of the wake around the streamwise velocity minimum locations occurs.  The locus of the velocity minima along the streamwise direction define a helical centerline of the meandering wake.  Over a period of 4 rotor rotations, the wake meandering convects downstream, and the amplitude grows.   A slightly different wake evolution is seen in Fig. \ref{fig:asn_inst}(b) for the Region 3 case.  Both the inner wake hub vortex core with helical precession and outer wake form but remain more columnar.  As discussed before, due to the blade pitch the outer wake is weaker compared to Region 2.  The outer wake remains columnar with slight distortions from interactions outside the wake while the inner wake slowly expands.  The onset of wake meandering does not occur until after $x/D=5$. \\
\indent Similar instantaneous snapshots of the streamwise velocity for the simulations for both turbine operating conditions without a nacelle are shown in Fig. \ref{fig:as_inst}.  Close to the turbine, a jet along the centerline is clearly evident in both simulations.  Instantaneously, the near wake region transitions into wake meandering near the same distance from the turbine.  \\
\indent Figure \ref{fig:exp_psd} shows comparisons of the measured and Fourier power spectral density of the streamwise component of the velocity fluctuations at the top tip position at several locations in the wake for both operating conditions. The non-dimensional frequency Strouhal number $St = f D/ U_{hub}$ is defined by the turbine diameter $D$ and hub height velocity $U_{hub}$.
 At $x/D = 2$, the measured spectra shows that there are peaks at $St = 2.3$ and $St = 0.8$ and a longer flat region in low frequency space.  The former peak is the rotor frequency and is also present in the simulation results.  Both signatures are muted in the Region 3 cases.  A lower frequency around $St \sim 0.2-0.3$ is present Fig. \ref{fig:exp_psd}(b), Fig. \ref{fig:exp_psd}(c), and Fig. \ref{fig:exp_psd}(d). The low frequencies are found in both measurements and simulations and are stronger in the simulations that include a nacelle model.  
Farther downstream the high frequencies in computed spectrum have slightly more energy possibly due to the LES not having sufficient grid resolution to adequately capture the high frequency modes.  However, the processes we are concerned about are the low frequency modes, which will be discussed below.  The energy in these frequencies is similar to the measured data.  \\
\indent Figure \ref{fig:exp_psd}(b) at $x/D=4$ shows the simulation with a nacelle model turbine operating Region 2 has more energy in the low frequency energy modes than the simulation without a nacelle.  Similarly, a difference in turbulent intensities in Fig. \ref{fig:turb_uu} is also present between the simulations with and without the nacelle model.  At $x/D=6$ and 9 locations in Fig. \ref{fig:exp_psd} the low frequency modes have similar energy consistent with the streamwise turbulence intensities in Fig. \ref{fig:turb_uu} at $x/D=5$ and 7 locations.   In the turbine operating Region 3, the energy for simulations with a nacelle model in the low frequency modes at locations $x/D=4,6$ and $9$ is higher than simulations without a nacelle and is consistent with streamwise turbulence intensity at the  $x/D=4, 5$ and 7 locations.  
The energy differences are another indication that without a nacelle the turbulent energy at low frequencies is reduced.  This further shows that the nacelle model is necessary for accurate simulations of a wind turbine. Without it, the high energy, low frequency contributions cannot be fully taken into account.   \\
\indent The contours on the $St$-$x$ plane of Fourier power spectral density of the streamwise component of the velocity fluctuations pre-multiplied by frequency $f\Phi_{11}$ for the simulations with the nacelle model are shown in Fig. \ref{fig:linefft} for three different radial locations.   This allows us to visualize the evolution of the frequency modes in the wake of the turbine.  Starting at the centerline $y_h = y-H=0$, there are several frequency regions of high energy.  Nearest to the turbines, $x/D=0.5$, a high energy mode is centered around $St=0.7$. Given the proximity to the turbine along the centerline, this is immediately recognized as the hub vortex.  The Strouhal number is similar to the measured hub vortex frequencies of other turbines \citep{iungo2013linear, howard2015statistics, viola2014prediction, foti2016wake}.   In the Region 2 case, the peak hub vortex energy occurs directly downstream of the turbine and quickly dissipates.  By $x/D=2$, the energy level has dropped significantly, but, relatively, it is the strongest energy mode in the flow at that axial location.  This indicates that the hub vortex contains most of the energy along the centerline.  However, after $x/D=2$ another, lower, frequency mode becomes present, the wake meandering frequency, most prevalent at $St=0.3$, which has been observed in numerous studies: $St=0.23$ \citep{okulov2007stability}, $St=0.28$ \citep{chamorro2013interaction}, $St = 0.15$ \citep{foti2016wake} and $St=0.15-0.25$ \citep{medici2008measurements}. Both the wake meandering frequency and the hub vortex frequency remain the dominant frequencies throughout the rest of the domain along the centerline, a clear indication that while the hub vortex breaks down, its remnants continue to affect the flow far downstream.  The centerline frequency contours of Region 3 have some drastic distinctions from the former case.  Here, the hub vortex energy peaks further from the rotor.  Between $2 < x/D < 4$, little low turbulent energy is present indicating the hub vortex has not broken down and no interaction with the outer wake has occurred.  By $x/D > 6$, much later than that of Region 2 and consistent with mean turbulent statistics, the wake meandering mode is activated.  Moreover, the interaction between the inner and outer wake not only affects the wake meandering but also strengthens the hub vortex frequency mode along the centerline very close to the turbine. \\
\indent In both cases near the mid-blade location, $y_h/D = 0.2$, the hub vortex frequency is not distinguishable from other frequency modes  close to the turbine.  Further downstream, the wake meandering frequency appears, first in Region 2 at $x/D=2.5$, consistent with the turbulence kinetic energy shown above.   The peak wake meandering energy occurs around $x/D=4$, and high energy is observed far downstream as the wake meanders throughout the far wake.  For Region 3 case, the wake meandering is observed to peak much further downstream, around $x/D=5$ and is also present far downstream. The energy present in the wake meandering mode at this radial position is higher compared to the centerline.  Here, there is more turbulence energy in the wake meandering mode in the wake because of the proximity to the location of the interaction of the expanding inner wake and outer wake.  Near the peak wake meandering location, higher frequency modes are also present and can be attributed to complex interaction of the inner wake including the hub vortex and outer wake.  \\
\indent Closer to the blade tips at $y_h/D = 0.44$, similar to the mid-blade location, the hub vortex frequency is not present near the turbine.  Higher frequencies associated with the rotor frequency and the tip vortices are present but quickly dissipate and become indistinguishable compared to wake meandering.  Consistent with the findings of the turbulence kinetic energy and the mid-blade power density spectrum, wake meandering in Region 2 begins earlier than Region 3.  The peak energy levels are slightly upstream compared with the mid-blade and centerline.  Turbulence energy in the outer wake is highest due to the strength of the tip shear layer and interaction with the hub vortex core formed behind the rotor.  From these contours, we conclude that wake meandering begins at the outer wake, and the interaction of the inner wake with the outer wake introduces more turbulence. Evidence from streamwise power spectra in Fig. \ref{fig:exp_psd} shows that without the expanding inner wake the energy of the wake meandering is diminished. From onset of the wake meandering in the outer wake shear layer, the meandering modes begins to propagate downstream and towards the centerline explaining why the centerline energy peak of wake meandering occurs slightly further downstream compared to the top tip location. \\
\subsection{Meander Profiles}\label{sec:meander}
\begin{figure}
   \begin{center}
      \includegraphics[width=\textwidth]{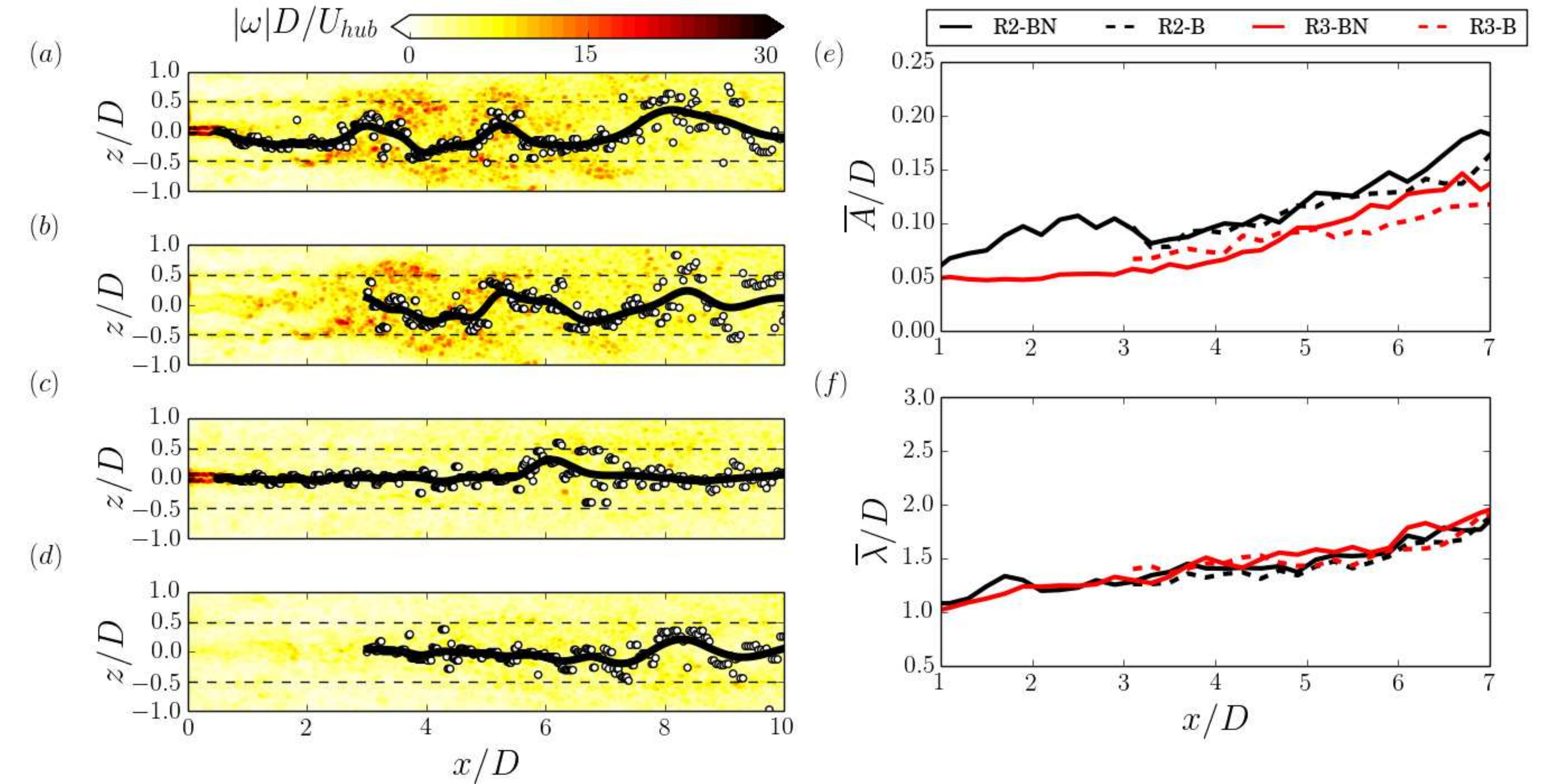}
      \caption{\label{fig:meander} Instantaneous vorticity magnitude, $|\omega|D/Uh$ with velocity minima (dot) and  meander profile (line) for (a) R2-BN, (b) R2-B, (c) R3-BN, (d) R3-B.  Average meander profile (e) amplitude, $\overline{A}$, and (f) wavelength $\overline{\lambda}$ with respect to distance from rotor plane, $x/D$, non-dimensionalized by the diameter $D$. }
   \end{center}
\end{figure}

\begin{figure}
   \begin{center}
      \includegraphics[width=\textwidth]{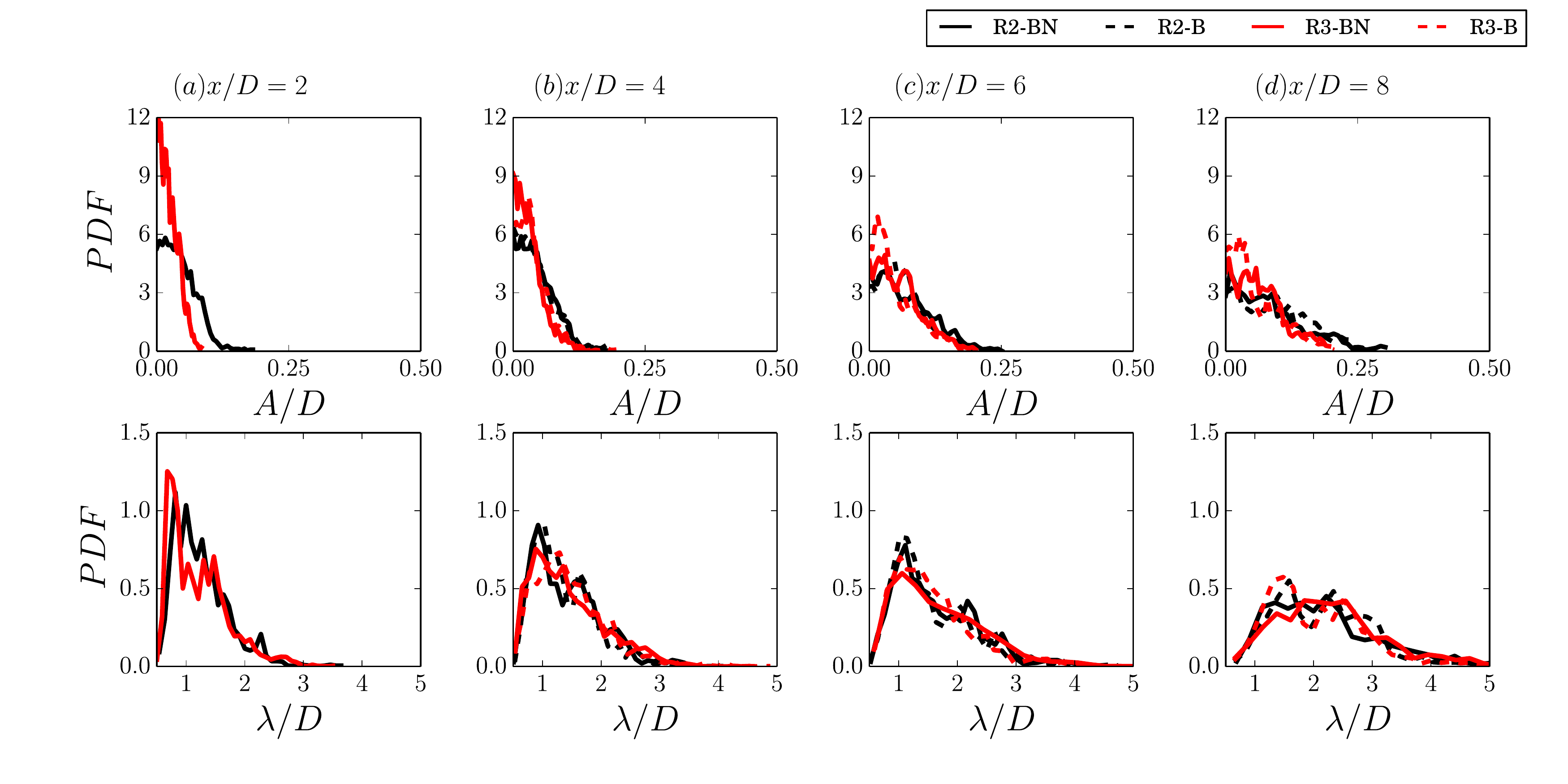}
      \caption{\label{fig:meander_pdf} Probability density function of the amplitude, $A/D$ (first row) and wavelength, $\lambda/D$ (second row) of the meander profiles at (a) $x/D=2$, (b) $x/D=4$, (c) $x/D=6$, and (d) $x/D=8$.}
   \end{center}
\end{figure}

\indent In this section, analysis is performed by reconstructing wake meandering into meander profiles that track the streamwise velocity minimum locations to further understand the dynamics of the wake of the turbine. In the instantaneous contours, Fig. \ref{fig:asn_inst}, we see that the streamwise velocity minima follow the center of the meandering wake.  By tracking these positions with a three-dimensional profile, we investigate the dynamics in terms of amplitude and wavelength of meanders and obtain statistics of the dynamical wake. The reconstruction technique was first described in \citet{howard2015statistics} and developed for temporal and spatial resolved three-dimensional LES flow field in \citet{foti2016wake}.  A three step procedure is utilized for the three-dimensional helical meandering profile reconstruction: i) Use finite temporal average scheme proposed by \citet{chrisohoides2003experimental} to find the coherent time scale $\tau_c$ of the wake, ii) Locate the streamwise velocity minima along the axial direction, iii) Spatially filter the velocity minimum locations to create a continuous profile.  \\
\indent The proposed method for reconstruction of the wake meandering is based on tracking wake meandering as a large-scale coherent structure in the wake of the turbine. The flow field is temporally filtered to eliminate the high frequency fluctuations which are not associated with the low frequency wake meandering.  The temporal filtering process uses finite averaging over a coherent time scale $\tau_c$ to decompose the flow field into the usual triple decomposition of a turbulent flow as described by \citet{hussain1986coherent}. A temporally and spatially evolving flow field $\boldsymbol{u}(\boldsymbol{x},t)$ (bold indicates a vector quantity) can be rewritten as follows: 
\begin{equation}
\boldsymbol{u}(\boldsymbol{x},t) = \boldsymbol{U}(\boldsymbol{x}) + \tilde{\boldsymbol{u}}(\boldsymbol{x},t) + \boldsymbol{u_i}(\boldsymbol{x},t), \label{eqn:triple}
\end{equation}
where $\boldsymbol{U}(\boldsymbol{x})$ is the mean, strictly spatial term, $\tilde{\boldsymbol{u}}(\boldsymbol{x},t)$ is the coherent term and $\boldsymbol{u_i}(\boldsymbol{x},t)$ is the incoherent term.  Based on the work of \citet{chrisohoides2003experimental}, the optimal size of an interval averaging window is equivalent to the coherent time scale and can be determined by starting with a finite averaging over a time window of $\tau$ as follows:
\begin{equation}
    u_\tau(x,t) = \frac{1}{\tau} \int^{t+\tau/2}_{t-\tau/2} u(x,t^\prime) dt^\prime.
    \label{eqn:interval}
\end{equation}
The proper coherence time scale can be found by employing fluctuation analysis and the central limit theorem.  Using a $\tau$ too small the large-scale motions will be dominated by incoherence, while averaging over too long will smear out the coherent structures.  Fluctuation analysis shows that the coherent time scale $\tau_c$ is the time window where the standard deviation over the temporal window begins to scale as $\tau^{-1/2}$. With this temporal window, the finite averaged velocity $u_\tau(x,t) = U(x) + \tilde{u}(x,t)$. For more background, see \citet{chrisohoides2003experimental} and \citet{foti2016wake}.
\indent The optimal coherent time scale for all cases is found to be about $\tau_c = 0.6T$.  This value is close to the optimal time scale of a model turbine discussed in \citet{foti2016wake}.  \\
\indent With the coherent time scale determined and the flow decomposed into its three parts, the minimum streamwise velocity, $U(x) + \tilde{u}(x,t)$, locations along the axial direction are tracked over 200 rotor revolutions at the coherent time scale resolution.  Physically, the velocity minima track near the center of the hub vortex near the turbine and meandering in the far wake.   A low-pass spatial filter using a length scale $l_c = D/2$ is used to obtain a three-dimensional continuous profile out from the velocity minima, as shown in \citet{howard2015statistics}. \\ 
\indent Figures \ref{fig:meander}(a)-(d) show examples of a three-dimensional meander profile projected on the hub height plane for both turbine operating conditions and simulations with and without the nacelle model.  In each figure, the solid line represents the meandering profile, and the circle markers are the velocity minimum locations in the wake.  The meander profile is superimposed on the instantaneous vorticity magnitude, $|\omega| D/U_{hub}$.  The trends addressed previously in the instantaneous streamwise snapshots in Fig. \ref{fig:asn_inst} about the dynamics of the wake for each case are observed in the meander profiles.  
In Fig. \ref{fig:meander}(a), the snapshot of the meander profile of the simulation with a nacelle model for the turbine operating in Region 2  shows an energetic wake and the meander profile that starts behind the nacelle and quickly expands tracking the hub vortex toward the outer wake.  Accompanying the expansion of the hub vortex and increasing amplitude of the meander profile is an increased vorticity magnitude near the peaks of the meander profile at the tip shear layer (shown by dashed lines on the Figures \ref{fig:meander}(a)-(d)). This is further evidence that the expansion of the hub vortex towards the tip shear layer occurs with increases in turbulence intensity as the tip shear layer and hub vortex interact as the meandering of the wake commences.  
Figure \ref{fig:meander}(b) shows the snapshot of the simulation without a nacelle model in Region 2.  The meander profile cannot start immediately downstream of the rotor because the centerline jet affects the location of the velocity minima near the turbine.  The meander profile starts at $x/D=3$ where the centerline jet dissipates. A meander profile of the Region 3 simulation with a nacelle is observed in Fig. \ref{fig:meander}(c).  Unlike Region 2 simulations, the inner wake is less energetic, and the meander profile is a tight spiral for 5 diameters behind the turbine. After $x/D=5$ the meander profile has large amplitudes towards the tip shear layer.  Similar to Region 2, instantaneous high vorticity regions are near the peaks of the meander profile.   The simulation without the nacelle model operating in Region 3 is shown in  Fig. \ref{fig:meander}(d), with similar trends as the simulation with the nacelle model. Further analysis of many meanders is necessary to understand the trends in the wake.  \\
\indent The statistics of the meander profiles are useful in determining the dynamics of the hub vortex and wake meandering.  
It is useful to investigate the wave-like characteristics, amplitude and wavelength, of the meander profiles as a function of axial distance from the turbine.  
The behavior of amplitude and wavelength can be interpreted in both the near wake and far wake separately.  
In order to obtain the wave features, the three-dimensional helical profile is projected onto the two-dimensional hub height plane where amplitudes $A$ and wavelengths $\lambda$ are readily calculated from the extrema of the profile. 
In Fig. \ref{fig:meander}(e), the average amplitude $\overline{A}/D$ as a function of the downstream distance for each test case is shown.  
In the simulation with a nacelle model in Region 2, the amplitude quickly increases behind the turbine with the expanding hub vortex.  
By $x/D=2$ the amplitude increases to a peak in the near wake followed by a plateau as hub vortex interacts with the outer wake.  The amplitude increases further as the wake begins to meander. 
On the other hand, in Region 3, the amplitude is much lower in the near wake, consistent with the tight spiraling hub vortex.
The amplitude increases linearly until $x/D = 5$ where the wake meandering commences. The average amplitude in Region 3 remains lower than the amplitude in Region 2 throughout the wake.    
Simulations without the nacelle yield meander profiles with lower amplitudes regardless of turbine operating condition.  In Region 2, the average amplitude between the simulations with and without the nacelle model remains similar until $x/D=4.5$ where the differences becomes more pronounced.  The amplitude is more attenuated for the simulation without a nacelle in Region 3 starting from $x/D=5$ where wake meandering is witnessed to commence.  We can relate the amplitude $A$ to the energy $E$ of the meander profile, $A^2 \sim E$. 
Consistent with the turbulence intensities in Fig. \ref{fig:turb_uu} and the power spectrum in Fig. \ref{fig:exp_psd}, the energy in the wake meandering is reduced by 40\% without a nacelle model.
Figure \ref{fig:meander}(e) provides quantitative evidence that a nacelle model is needed to adequately simulate the dynamics of the wake. \\
\indent The average wavelength $\overline{\lambda}/D$ of the meander profiles is shown in Fig. \ref{fig:meander}(f).  Unlike the amplitudes for each case, the average wavelength is approximately the same regardless of the nacelle model. The similarity of the profiles is indicative that the wavelength is a large-scale feature of the flow not pertaining to the dynamics.  For all cases in the far wake, the mean recovery of the wake is generally the same.  The evidence shows that the mean elongation of the meandering profile is caused by the wake of the turbine.  As the wake recovers, the meander is stretched in the streamwise direction. \\
\indent The statistics of the meander profiles are further investigated in Fig. \ref{fig:meander_pdf} with the probability density function (PDF) of the amplitude and the wavelength shown at different locations downstream.  
The amplitude PDF for the simulation with a nacelle model in Region 2 flattens as the location from the turbine increases.  At $x/D=2$ the highest amplitudes extends only $A/D = 0.15$. The maximum amplitudes eventually extend beyond $A/D=0.25$, indicating that the centerline of the wake is being displaced out to the tip shear layer by wake meandering.  Conversely, the amplitude PDF of the simulation with a nacelle model in Region 3 has both a lower median and standard deviation than Region 2.  The standard deviation of the amplitude normalized by the average amplitude $\sigma_A/\overline{A}$, an indication of the uncertainty in the amplitude, is nearly constant throughout the domain and ina ll simulations around 0.7.   
At $x/D=2$ location, the PDF is very thin, consistent with the tightly spirally hub vortex.  At locations further downstream, the amplitudes increase but the maximum in Region 3 is always less than Region 2.  With regards to the simulations without a nacelle mode, they are not shown at $x/D=2$ due to the manifestation of the unphysical centerline jet which affects the location of the velocity minima in the inner wake. At locations further downstream, the simulations without the nacelle have slightly lower amplitude medians than the corresponding simulation with nacelle model.  The maximum amplitudes achieved are also lower.  \\
\indent The PDFs of the wavelength show a gradual growth of the median and variance as the locations from the turbine get further downstream.  At $x/D=4$, most wavelengths for all cases are near $\lambda/D=1$.  By $x/D=8$, the wavelengths increase significantly and the PDFs all have a flattened peaks from $1 < \lambda/D < 3$.  Although the wavelength PDFs of simulations with a nacelle model far from the turbine have a slight shift towards higher wavelengths, the wavelength for both operating conditions and nacelle model are not substantially different. \\

%
%
\subsection{Dynamic Mode Decomposition and Meander Profile}\label{sec:dmd}
\begin{figure}
   \begin{center}
      \includegraphics[width=\textwidth]{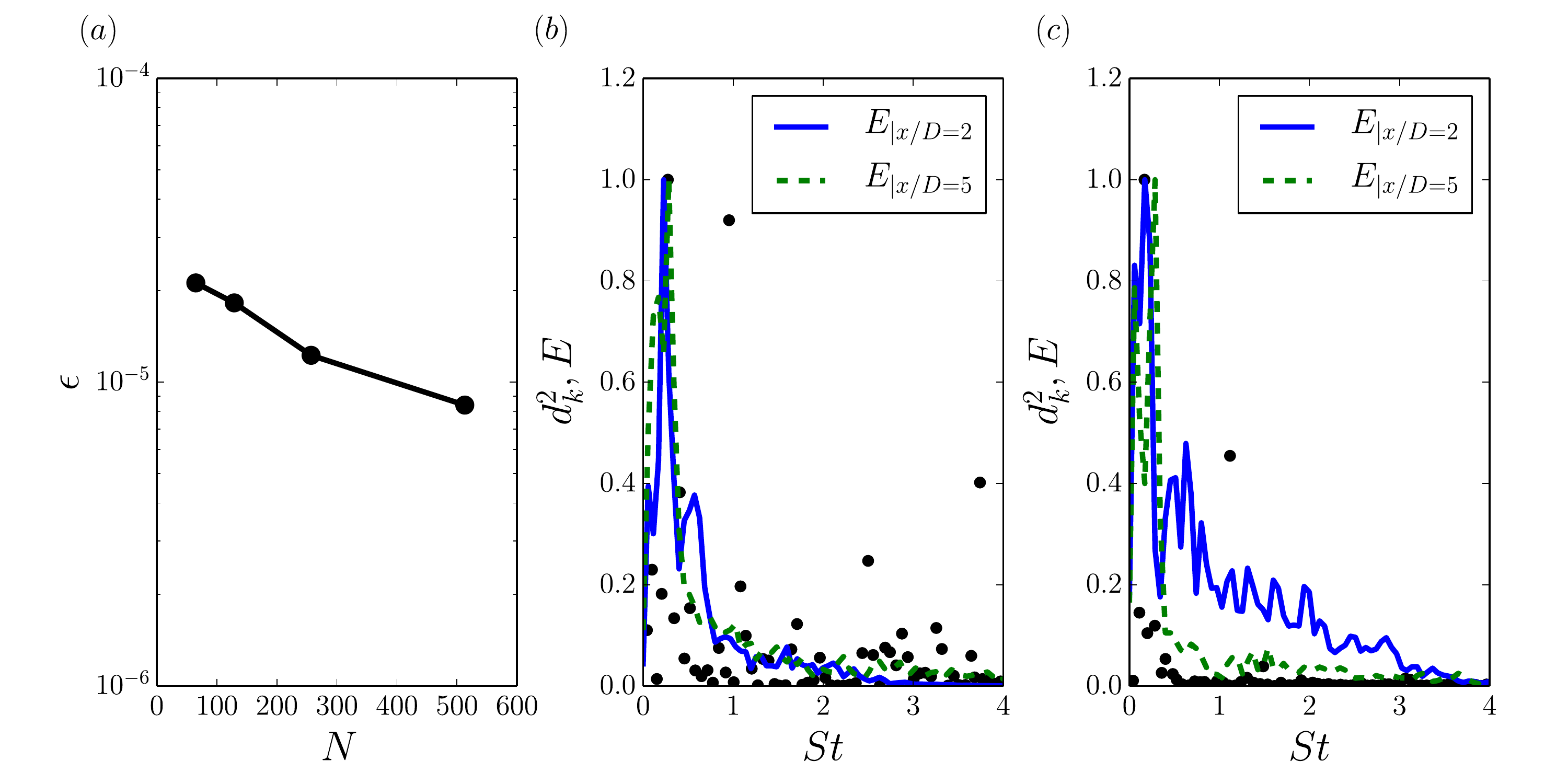}
      \caption{\label{fig:dmd_validation} Validation of dynamic mode decomposition. (a) Residuals norm $\epsilon$ of DMD modes as number of modes $N$ increases. Dynamic mode energy spectrum $d_k^2$ (circle), energy spectrum at $x/D=2$ (solid), energy spectrum at $x/D=5$ (dashed) for (b) R2-BN and (c) R3-BN.}
   \end{center}
\end{figure}

\begin{figure}
   \begin{center}
      \includegraphics[width=\textwidth]{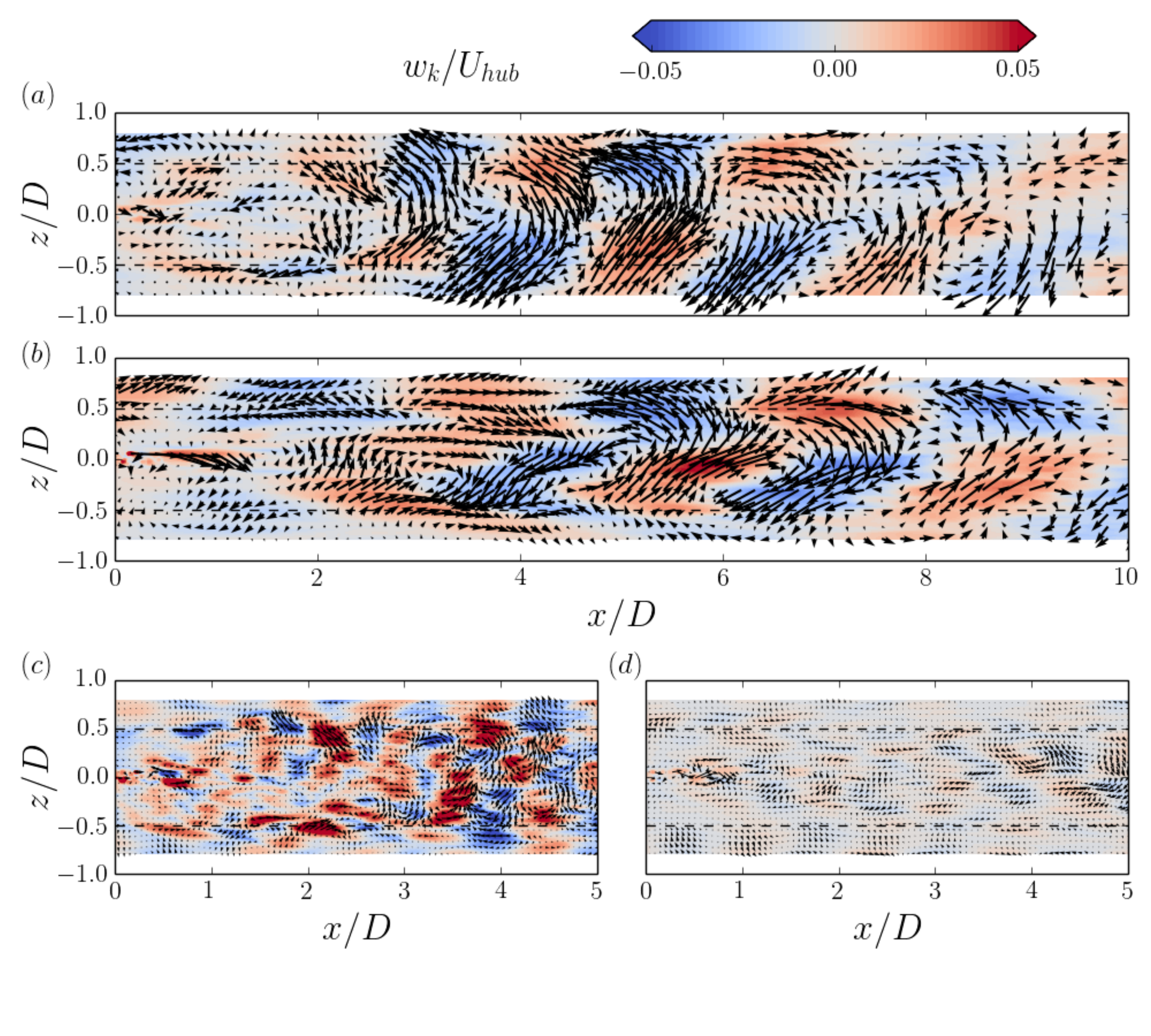}
      \caption{\label{fig:dmd_modes}  Selected dynamic modes visualized by contours of the streamwise velocity and two dimensional ($x-z$) plane vector field. (a) R2-BN: $St=0.29$, (b) R3-BN: $St=0.3$, (c) R2-BN: $St=0.73$, and (d) R3-BN: $St=0.74$.}
   \end{center}
\end{figure}

\begin{figure}
   \begin{center}
      \includegraphics[width=\textwidth]{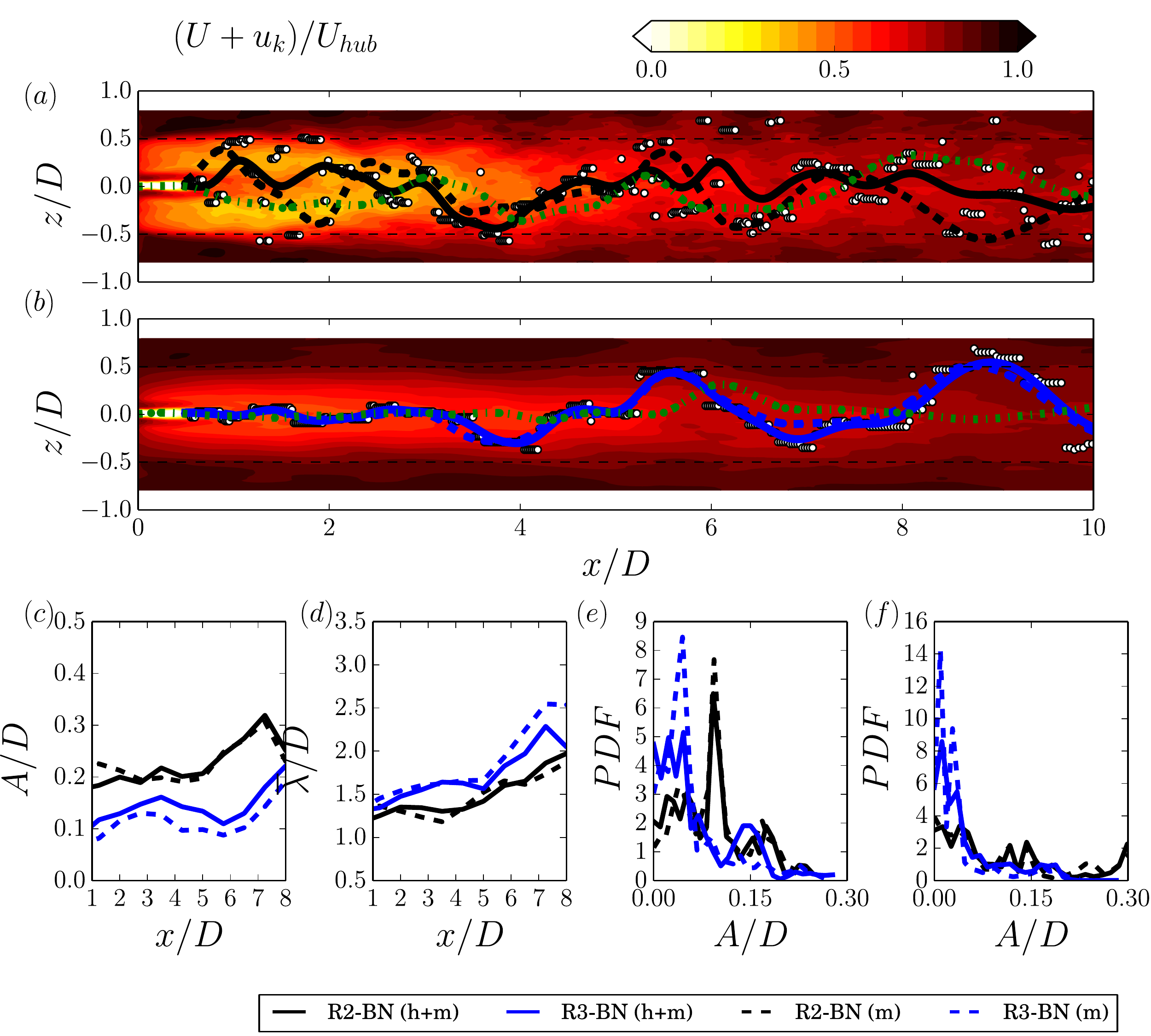}
      \caption{\label{fig:dmd_meander} Dynamic mode meander profiles created from hub vortex and wake meandering mode (h+m), dynamic mode meander profile from wake meandering mode (m) and meander profile from complete flow field overlaid on contours and velocity minima (white dots) of the sum of the average velocity and selected hub vortex and wake meandering modes, $U + u_k$, non-dimensionalized by the hub velocity $U_{hub}$ for (a) R2-BN and (b) R3-BN. For reference, the corresponding temporal averaged wake meander profile is shown (dashed-dotted line). Characteristics of dynamic mode meander profiles (c) amplitude, $\overline{A}$, and (d) wavelength $\overline{\lambda}$ with respect to distance from rotor plane. Probability density function of amplitude, $A/D$, at (e) $x/D=3$ and (f) $x/D=6$.}
   \end{center}
\end{figure}

\begin{figure}
   \begin{center}
      \includegraphics[width=\textwidth]{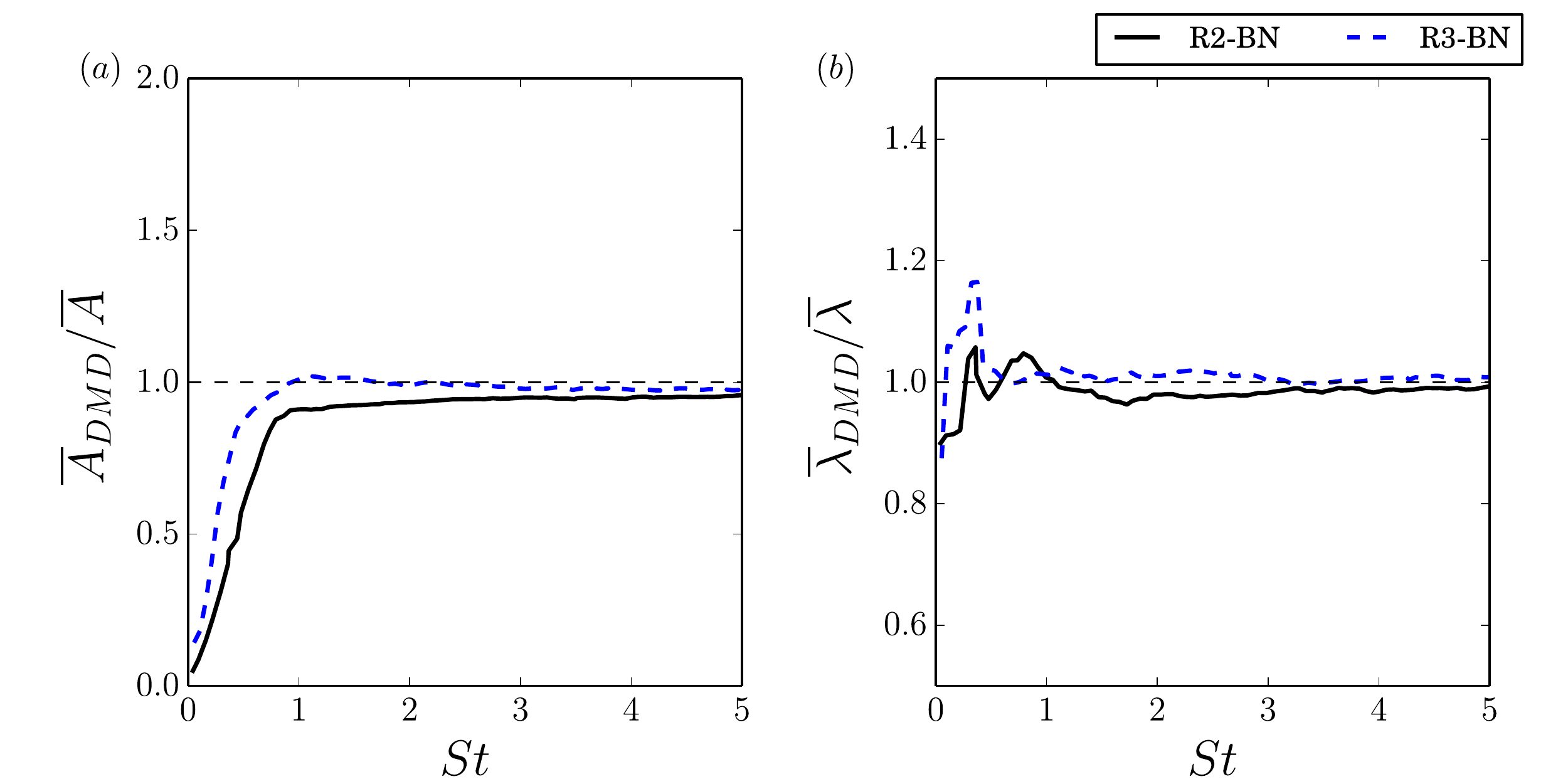}
      \caption{\label{fig:dmd_meander_ratio} Ratio of (a) the averaged amplitude and (b) the averaged wavelength of the dynamic mode meander profiles to those from the complete flow field as the number of employed frequency modes in ascending order are increased at $x/D=8$.}
   \end{center}
\end{figure}

\indent Up to this point, we have concentrated the development of two large coherent structures in the wake: the hub vortex and wake meandering.  Each structure is associated with a dominant frequency, and the dynamics is interpreted through the meander profiles reconstruction.  To further elucidate these coherent structures and their effect on the wake of the turbine, we decompose the wake using dynamic mode decomposition (DMD).  DMD was first theoretically introduced by \citet{rowley2009spectral} as a technique to decompose the flow by spectral analysis of the Koopman operator, a linear operator associated with the full non-linear system.  The Arnoldi-like method was improved by \citet{schmid2010dynamic} and is able to compute the modes of a finite sequence of snapshots of the flow field.  \citet{sarmast2014mutual} used DMD to separate wind turbine tip vortex modes for analysis. The modal decomposition is able to extract spatial structures and their corresponding frequencies and growth rates without having the explicit dynamic operator.  Because DMD can separate specific spatial modes by their individual frequencies, we can explicitly extract the modes related to the hub vortex and wake meandering.  With the spectral analysis performed above, the specific frequencies of the hub vortex and meandering wake are readily known from Fig. \ref{fig:linefft}. \\
\indent First, we will give a brief overview of the algorithm.  For more information, please see \citet{schmid2010dynamic} and \citet{sarmast2014mutual}.  For our simulations, a sequence of three-dimensional instantaneous flow fields $\boldsymbol{u}_i(\boldsymbol{x}_j,t_i), t_i = i\Delta t, i=0,1,N-1$ are assembled into a matrix
\begin{equation}
   U_n = [\boldsymbol{u}_0, \boldsymbol{u}_1,..., \boldsymbol{u}_{N-1}],
\end{equation}
where $N$ is the number of snapshots, and $\Delta t$ is the time between each snapshot.  In DMD, a linear mapping $A$ or $\widetilde{A}$ is assumed to relate a flow field $\boldsymbol{u}_j$ to the succeeding flow field $\boldsymbol{u}_{j+1}$ such that
\begin{equation}
   \boldsymbol{u}_{j+1} = A \boldsymbol{u}_j = e^{\widetilde{A}\Delta t}\boldsymbol{u}_j,
\end{equation}
and decomposing the flow field into spatial eigenmodes $\phi_k(\boldsymbol{x}_j)$ and temporal coefficients $a_k(t_i)$
\begin{equation}
   \boldsymbol{u}_{j+1} = \sum_{k=0}^{N-1} \boldsymbol{\phi}_k a_k = \sum_{k=0}^{N-1} \boldsymbol{\phi}_k e^{\imath \omega_k \Delta t} = \sum_{k=0}^{N-1} \boldsymbol{\phi}_k \lambda_k^j
\end{equation}
where the $\imath \omega_k$ is the corresponding eigenvalues of $\widetilde{A}$, and $\lambda_k^j$ is the corresponding eigenvalues of $A$.  The amplitude $d_k$ and energy $d_k^2$ of spatial dynamic mode $\boldsymbol{\phi}_k$ are related such that $\boldsymbol{\phi}_k = d_k \boldsymbol{v}_k$ where $\boldsymbol{v}_k^T \boldsymbol{v}_k = 1$. The eigenvalues $\lambda_k$, also referred to as the Ritz values, are complex conjugates which all lie on the complex unit circle, $|\lambda_k| = 1$. To obtain the more familiar complex frequency $\imath \omega_k = \log(\lambda_k) / \Delta t$. The real part is the temporal frequency, and the imaginary part is an exponential growth rate of the dynamic mode.\\
\indent A substantial number of three-dimensional instantaneous snapshots are saved for each simulation in order to cover the amount of time needed to resolve the low frequencies of the hub vortex and wake meandering.   Each snapshot contains the computational cells that are within a diameter wide and high box centered on the turbine and entire length of the computational domain.  The mean flow is subtracted from each snapshot to obtain the  fluctuating part of the flow field. The minimum time between the snapshots $\Delta t = T/12$, where $T$ is the rotor period, a time difference low enough to ensure that not only the low frequency modes but the blade rotation frequency are captured as well. To ensure convergence of mode decomposition, the number of snapshots used are increased until the norm of the residuals, $\epsilon$, of the mapping operator becomes sufficiently small.  Figure \ref{fig:dmd_validation}(a) shows the residuals decrease by several orders of magnitude as the number of snapshots is increased to $N = 512$.  Based on the residuals, the number of snapshots used for subsequent analysis will be $N=512$.  Moreover, several series of snapshots are used and averaged so the modes and frequencies are averaged over a time spanning $100T$. With a $\Delta t = T/12$ and $N=512$, the low frequency precession of wake meandering will occur as many as five times, enabling its temporal resolution for our proposes. \\
\indent The energy $d_k^2$ and its frequency $St = \Re( 2 \pi \imath \omega_k ) D/U_{hub}$ of $k$th mode for simulations with the nacelle model in Region 2 and Region 3 is shown in Fig. \ref{fig:dmd_validation}(b) and Fig. \ref{fig:dmd_validation}(c), respectively.  The energy of each mode is normalized by the maximum energy.  The maximum energy is associated with the wake meandering frequency $St=0.3$. Also included in each figure are two normalized energy spectra: one located in the near wake at $x/D=2$ and one located the far wake at $x/D=5$.  Both energy spectra have a maximum energy at $St=0.3$ (Note both the extrema of energy of dynamic modes and the energy spectra occur at the same frequency). A few energy peaks are present in the DMD.  In the low frequency region, $St<1$, the dynamic mode energy peaks at the aforementioned $St=0.3$ and $St=0.74$, similar frequencies to what was determined to be the wake meandering frequency and hub vortex frequency, respectively.  The energy spectra confirm that DMD decomposes the flow field into modes corresponding to modes present in the spectral analysis with hub vortex frequency is only present in the energy spectrum at $x/D=2$ and the wake meandering is present at both energy spectra shown.  However, due to the location chosen for the energy spectra, the high frequencies related to the rotor frequency are not captured but are found readily in DMD. In the higher frequencies of the dynamic mode, $St > 1$, there are several peaks including the rotor frequency of $St=2.3$.  The energized modes with high frequencies including $St=1$ relate to the rotor frequencies. Most high frequencies have a lower contribution to the energy of the flow compared to the low frequency modes like wake meandering and are related to energy modes present very close to the turbine blades. \\
\indent Next, we begin analysis on the spatial modes provided by DMD of the simulations with the nacelle model for both operating conditions. We select the modes pertaining to the meandering wake and the hub vortex.  Figure \ref{fig:dmd_modes}(a) shows the wake meandering dynamic mode with a $St = 0.29$ represented by the streamwise coherent velocity $u_k$ and the two-dimensional ($x-z$) plane vector field for the simulation in Region 2.  The dynamic mode shows structures of sources and sinks of the streamwise coherent velocity at $x/D>2$ consistent instantaneous velocity of wake meandering from Fig. \ref{fig:asn_inst}(a).  The structures expand in the streamwise direction similar to wake meandering shown through the wavelength elongation due to the recovering wake.  The streamwise coherent velocity from the dynamic mode corresponding to wake meandering with a frequency $St = 0.3$ in the simulation in Region 3 is shown in Fig. \ref{fig:dmd_modes}(b).  
Similar to the dynamic mode of the simulation in Region 2, sinks and sources in the coherent velocity appear downstream of the turbine.  The wake meandering features begin to form further downstream in agreement with the simulation with the nacelle model in Region 3 mean and instantaneous flow fields.  The meandering patterns are slightly weaker and are pulled more towards the centerline than in the simulation in Region 2.  Another noteworthy feature the wake meandering dynamic mode in both simulations is the prominent velocity regions around location of the hub vortex.  The velocity field in the mode shows an elongation and expansion of the hub vortex.  
Separate spatial dynamic modes contains more of the dynamics of the hub vortex. The strongest hub vortex mode in the simulation in Region 2 with a frequency $St=0.73$, with an energy $d^2_k$ about 40\% of the wake meandering energy, is shown in Fig. \ref{fig:dmd_modes}(c).  There is a strong streamwise velocity near the nacelle, indicative of the meandering in the inner wake caused by the hub vortex. The streamwise velocity expands outwards towards the tip shear layers quickly in agreement with the analysis of the instantaneous wake.  Downstream of the expansion towards the tip shear layers, the coherent meandering of the hub vortex is lost, but high fluctuations persists far downstream. The mode demonstrates that the hub vortex has an impact downstream as spectral analysis in Fig. \ref{fig:linefft}(a) suggests.  
Figure \ref{fig:dmd_modes}(d) shows the streamwise coherent velocity dynamic mode of the simulation in Region 3 with a frequency of $St=0.74$. The dynamic mode of the hub vortex is very different than the simulation from Region 2. The streamwise velocity remains in its tight spiral around the centerline with weak positive and negative regions of the streamwise coherent velocity.  The Region 3 hub vortex mode is further evidence that the hub vortex does not expand quickly and interact with the tip shear layer with a similar intensity as seen in the simulation from Region 2. It is clear that operating condition affects the expansion and stability of the hub vortex. \\
\indent To elucidate and characterize how certain frequencies create the dynamics in the far wake, the dynamic modes related to wake meandering and hub vortex shown in Fig. \ref{fig:dmd_modes} are extracted to obtain a coherent velocity and used in conjunction with the meander reconstruction analysis described in the previous section \ref{sec:meander}.  
Instead of using the finite temporal averaging to obtain a coherent velocity, the coherent velocity of the selected dynamic modes is used, and the velocity is used to create dynamic mode meander profiles with similarities to the meander profile statistics shown above.
Two dynamic mode meandering profiles are created from selected dynamic modes: i) Summing the wake meandering mode and hub vortex mode (h+m) and ii) Selecting only the wake meandering mode (m).
The mean velocity and coherent velocity $U + u_k = U + \widetilde{u}$  are used to find where the velocity minima are located at each axial location downstream of the turbine just like described in the section above.  The dynamic mode meander profile is obtained by low-pass spatial filtering of the velocity minima for a continuous profile.  Dynamic mode meander profiles are collected for each time instance using the reconstructed velocity from only selected dynamic modes. \\
\indent Figure \ref{fig:dmd_meander}(a) shows an instance of both dynamic mode meander profiles and with a meander profile created using finite average filtering overlaid on the velocity $U + u_k$ from the simulation with the nacelle model in Region 2. 
Immediately downstream of the turbine, both dynamic mode meander profiles do have an increased amplitude but a similar wavelength to the finite average filtering.  In the near wake, the meander profiles created with and without using the hub vortex mode show significant difference until $x/D=2.5$ where the profiles begin to converge.  The hub vortex mode interacts with the wake meandering mode.  However, the hub vortex mode has a large contribution to the meander profile after $x/D=5$ where both profiles begin to diverge again. 
The dynamic meander profiles from the simulation in Region 3, shown in Fig. \ref{fig:dmd_meander}(b), are similar to each other until $x/D=6$ where the profiles begin to diverge.  This is in contrast to the Region 2 profile because in Region 3 the hub vortex mode is significantly weaker. \\
\indent The average amplitude $\overline{A}/D$ of the dynamic mode meander profiles is shown in Fig. \ref{fig:dmd_meander}(c).  The statistics of the dynamic mode meander profile for Region 2 reveal that the mean amplitude is slightly higher for the (h+m) profile which includes the hub vortex compared to the (m) profile. There is not a significant difference.  The hub vortex mode has a large effect on the average amplitude of the dynamic mode meander profile for simulation in Region 3.  The average amplitude for the (h+m) profile is significantly higher.  
The average amplitude for both turbine operating regions is able to capture the trends of the wake meandering and is higher compared with the corresponding meander profile amplitude shown in Fig. \ref{fig:meander}(e).  The higher amplitude in the far wake for the dynamic modes suggests that there are some higher frequency modes that smooth wake meandering.
However, the wavelength statistics in Fig. \ref{fig:dmd_meander}(d) show conclusively that the two dynamic modes can fully capture the expansion and elongation of the meandering wake as the average wavelength compares well with the Fig. \ref{fig:meander}(f).  The wavelength grows relatively linearly, and it suggests that the wake meandering dynamic mode is responsible because of the gradual elongation of the coherent regions in the streamwise velocity.  
Probability density functions of the amplitudes of the dynamic mode meander profiles at $x/D=3$ and $x/D=6$ are shown in Fig. \ref{fig:dmd_meander}(e) and Fig. \ref{fig:dmd_meander}(f), respectively. Noticeable is that the meander profiles that contain the hub vortex has a higher probability of larger amplitudes.  While using two modes should affect the amount of energy, compared to only using the wake meandering mode, the energy and amplitude far from the turbine is higher where the hub vortex mode must have a large effect. \\ 
\indent The above analysis shows that at minimum two dynamic modes are needed to capture most of the fundamental features of wake meandering downstream in the wake. While the energy of the meandering wake is slightly over-predicted with the two modes, the wavelength is captured quite well.  Figure \ref{fig:dmd_meander_ratio}(a) shows the ratio of the average amplitude of the dynamic mode meander profile to the filtered meander profile of the complete flow field at $x/D=8$ as the number of dynamic modes, arranged by frequency in ascending order, are included in the flow field.  It shows conclusively that amplitude of the meandering profile using dynamic mode decomposition converges to the meander profile using the temporal filtering with just a few of the low frequency modes for either operating condition.  Both profiles from each operating condition increase monotonically from $St=0$ to $St = 0.74$, the hub vortex frequency.  After, the ratio slowly converges to unity. The low frequencies are the most important in capturing the meandering wake. This is a further indication that the low frequency meandering is captured using a few modes, namely, the modes corresponding to the wake meandering and hub vortex frequency. Figure \ref{fig:dmd_meander_ratio}(b) similarly shows the ratio of the average meander profile wavelength of the dynamic mode decomposition to temporal filtering meandering profile at $x/D=8$.  The wavelength ratio peaks for both operating conditions at $St = 0.34$, near the wake meandering frequency, but when the hub vortex frequency is included the ratio trends towards unity showing that both modes are necessary for prescribing the dynamic motions of the wake.  
%
%
%
%
%
\section{Discussion and Conclusions}\label{sec:conclusion}
\indent Wind tunnel experiments and large-eddy simulations are carried out to investigate the flow past a model wind turbine with a rotor diameter of 1.1 meters sited in a wind tunnel operating in Region 2 and Region 3.  
The incoming turbulent flow is generated by spires at the inlet of wind tunnel in both the experiments and simulations.  The flow field statistics of the wake behind the wind turbine predicted by the simulations are in good agreement with measurements.
Simulations with and without a nacelle model are carried out to systematically study its effect on the hub vortex and meandering motions at far wake locations.  
While the simulations without a nacelle model accurately predict the mean velocity profiles in the far wake, they fail to capture the wake just behind the nacelle, instead predicting a jet and significantly different distributions of turbulence intensity close to the turbine along the centerline.  Specifically, the hub vortex along the centerline remains a columnar jet for the simulations without a nacelle model.   Wake meandering with its downwind locations well correlate with the high turbulence intensity region for both simulations with and without a nacelle model. However, the simulations without a nacelle model predict lower turbulence intensity at far wake locations, and the location for maximum turbulence intensity is further downwind. Significant influence of turbine operating conditions on far wake turbulence intensity and wake meandering is observed: large amplitude wake meandering happens further downwind and the length of the region with high turbulence intensity is significantly longer for the Region 3 operating condition.\\
\indent   To further investigate the characteristics of wake meandering for different operating conditions and the effects of the nacelle, the instantaneous large-scale coherent wake meander profiles are constructed using the finite time averaging technique developed by \citet{howard2015statistics} and \citet{foti2016wake}.  We show that the amplitude of wake meandering is less for the simulations without a nacelle model for both operating conditions.  Also, the amplitude is lower in the turbine operating in Region 3 in comparison to Region 2. However, the wavelength and growth rate of wavelength for wake meandering are similar at different downwind locations regardless of nacelle modelling or turbine operating conditions.   \\ 
\indent The  wake meandering frequency is associated with a $St \sim 0.3$ in many different studies (\citet{medici2008measurements, chamorro2013interaction, okulov2014regular}), while the hub vortex is measured to have a frequency with $St \sim 0.7$ in \citet{iungo2013linear}, \citet{howard2015statistics} and  \citet{foti2016wake}.   For the present turbine, frequencies in those ranges are found for both turbine operating conditions.  The spectral analysis reveals that the presence of these frequencies correlated with the locations where the coherent motions of wake meandering and the hub vortex are most prevalent.  Dynamic mode decomposition of the flow allows us to extract and isolate the select modes related to both the wake meandering frequency and the hub vortex frequency.  The wake meandering dynamic mode consists of the large coherent oscillatory motions in the far wake.  The hub vortex dynamic mode, while strongest around the nacelle, persists far downstream, especially in Region 2 where the hub vortex is more unstable.  Two meandering profiles are created, using only the wake meandering mode and using both the wake meandering and hub vortex modes, respectively.  It is observed that the amplitude of the meandering profile using both modes is higher than that only using the wake meandering mode, especially for downstream locations.  \\
\indent From the evidence provided through our analyses, our work has further confirmed the importance of the nacelle on wake meandering at the far wake for a 1.1 meters diameter wind turbine under different operating conditions in addition to previous works for a 0.5 meters diameter hydrokinetic turbine \citep{kang2014onset} and a 0.13 meters model wind turbine \citep{foti2016wake}.  Future work will look at the nacelle effect for utility scale wind turbines and investigate similarity between scales for the hub vortex and wake meandering.
\begin{acknowledgements}  
This work was supported by U.S. Department of Energy (DE-EE0002980, DE-EE0005482 and DE-AC04-94AL85000), Xcel Energy through the Renewable Development Fund (grant RD4-13), and Sandia National Laboratories. Computational resources were provided by Sandia National Laboratories and the University of Minnesota Supercomputing Institute. Sandia National Laboratories is a multimission laboratory managed and operated
by National Technology and Engineering Solutions of Sandia, LLC, a wholly owned subsidiary of Honeywell International, Inc., for the U.S. Department of Energy’s National Nuclear Security Administration under contract DE-NA0003525.
\end{acknowledgements}
\bibliography{./biblio}

\end{document}